\documentclass[onecolumn]{aastex631}

\makeatletter

\newcommand{\Rmnum}[1]{\expandafter\@slowromancap\romannumeral #1@}

\graphicspath{{./}{figures/}}
\usepackage{graphicx}
\usepackage{appendix}
\usepackage{CJKutf8}
\usepackage{longtable} 
\usepackage{booktabs} 
\usepackage{comment}
\begin{document}
\begin{CJK}{UTF8}{gbsn}

\title{SN 2021dbg: A Luminous Type IIP-IIL Supernova Exploding from a Massive Star with a Layered Shell}
\correspondingauthor{Jujia Zhang}
\email{jujia@ynao.ac.cn}
\author{Zeyi Zhao}
\affiliation{Yunnan Observatories (YNAO), Chinese Academy of Sciences (CAS), Kunming 650216, China}
\affiliation{School of Astronomy and Space Science, University of Chinese Academy of Sciences, Beijing, 101408, China}
\affiliation{International Centre of Supernovae, Yunnan Key Laboratory, Kunming,
650216, China}
\affiliation{Key Laboratory for the Structure and Evolution of Celestial Objects, CAS, Kunming, 650216, China}

\author{Jujia Zhang}
\affiliation{Yunnan Observatories (YNAO), Chinese Academy of Sciences (CAS), Kunming 650216, China}
\affiliation{International Centre of Supernovae, Yunnan Key Laboratory, Kunming,
650216, China}
\affiliation{Key Laboratory for the Structure and Evolution of Celestial Objects, CAS, Kunming, 650216, China}

\author{Liping Li}
\affiliation{Yunnan Observatories (YNAO), Chinese Academy of Sciences (CAS), Kunming 650216, China}
\affiliation{International Centre of Supernovae, Yunnan Key Laboratory, Kunming,
650216, China}
\affiliation{Key Laboratory for the Structure and Evolution of Celestial Objects, CAS, Kunming, 650216, China}

\author{Qian Zhai}
\affiliation{Yunnan Observatories (YNAO), Chinese Academy of Sciences (CAS), Kunming 650216, China}
\affiliation{Key Laboratory for the Structure and Evolution of Celestial Objects, CAS, Kunming, 650216, China}

\author{Yongzhi Cai}
\affiliation{Yunnan Observatories (YNAO), Chinese Academy of Sciences (CAS), Kunming 650216, China}
\affiliation{Key Laboratory for the Structure and Evolution of Celestial Objects, CAS, Kunming, 650216, China}

\author{Shubham Srivastav}
\affiliation{Astrophysics Research Centre, School of Mathematics and Physics, Queen’s University Belfast, Belfast BT7 1NN, UK}

\author{Xiaofeng Wang}
\affiliation{Department of Physics, Tsinghua University, Beijing, 100084, China}
\affiliation{Purple Mountain Observatory, Chinese Academy of Sciences, Nanjing, 210023, China}

\author{Han Lin}
\affiliation{Yunnan Observatories (YNAO), Chinese Academy of Sciences (CAS), Kunming 650216, China}
\affiliation{International Centre of Supernovae, Yunnan Key Laboratory, Kunming,
650216, China}
\affiliation{Key Laboratory of Radio Astronomy and Technology, CAS, A20 Datun Road, Chaoyang District, Beijing, 100101, China}

\author{Yi Yang}
\affiliation{Department of Astronomy, University of California, Berkeley, CA 94720-3411, USA}
\affiliation{Department of Physics, Tsinghua University, Beijing, 100084, China}

\author{Alexei V. Filippenko}
\affiliation{Department of Astronomy, University of California, Berkeley, CA 94720-3411, USA}

\author{Thomas G. Brink}
\affiliation{Department of Astronomy, University of California, Berkeley, CA 94720-3411, USA}

\author{WeiKang Zheng}
\affiliation{Department of Astronomy, University of California, Berkeley, CA 94720-3411, USA}

\begin{abstract}
We present extensive observations and analysis of supernova (SN) 2021dbg, utilizing optical photometry and spectroscopy. For approximately 385 days following the explosion, SN 2021dbg exhibited remarkable luminosity, surpassing most SNe II. This initial high luminosity is potentially attributed to the interaction between the ejected material and the surrounding circumstellar material (CSM), as evidenced by the pronounced interaction signatures observed in its spectra. The subsequent high luminosity is primarily due to the significant $^{56}$Ni ($0.17 \pm 0.05$\,M$_{\odot}$) produced in the explosion. Based on the flux of flash emission lines detected in the initial spectra, we estimate that the CSM mass near the progenitor amounted to $\sim$(1.0--2.0) $\times 10^{-3}$\,M$_{\odot}$, likely resulting from intense stellar wind activity 2--3\,yr preceding the explosion. Considering the bolometric light curve, nebular spectrum modeling, and mass-loss rate, we suggest that the progenitor of SN 2021dbg was a red supergiant (RSG) with a mass of $\sim 20$\,M$_{\odot}$ and a radius of 1200\,R$_{\odot}$. This RSG featured a thick hydrogen shell, which may have contained a region with a sharp decrease in material density, electron density, and temperature, contributing to its layered structure. This object demonstrates mixed features of SNe IIP and SNe IIL, making it as a transitional event linking the above two subclasses of SNe II. 
\end{abstract}


\section{Introduction} 

Type \Rmnum{2} supernovae (SNe \Rmnum{2}) come from the explosion caused by the gravitational collapse of the dense iron core of hydrogen-rich massive stars at the end of thermonuclear evolution. According to the characteristics of light curves and spectra, they can be divided into SNe \Rmnum{2}P, SNe \Rmnum{2}L, SNe \Rmnum{2}n, and SNe \Rmnum{2}b \citep[e.g.,][]{1997ARA&A..35..309F,2012ApJ...756L..30A,2014Natur.509..471G,2017hsn..book..195G,2017hsn..book.....A}. In a volume-limited sample of the contemporary Universe, SNe \Rmnum{2}P account for about half of all core-collapse SNe \citep{2017suex.book.....B}, characterized by a plateau in the light curve lasting for about 3--4 months, followed by a rapid decline to a slower tail phase. The luminosity of the plateau is thought to result from the thermalization of the initial shock wave and the recombination of ionized hydrogen \citep[e.g.,][]{1993ApJ...414..712P}. SNe \Rmnum{2}L do not have the same obvious light-curve plateau characteristics as SNe \Rmnum{2}P, and the apparent magnitude decreases linearly from the peak -- that is, exponential decay of the luminosity. The mean maximum brightness of SNe \Rmnum{2}L is $\sim 1.5$\,mag brighter than that of SNe \Rmnum{2}P \citep{1993A&AS...98..443P,1994A&A...282..731P,2011MNRAS.412.1441L,2014ApJ...786...67A,2014MNRAS.442..844F,2014MNRAS.445..554F,2015ApJ...799..208S}, and limited evidence from archival images and SN rates suggests that the zero age main sequence (ZAMS) mass of SNe \Rmnum{2}L progenitors tends to be $\sim 20$\,M$_{\odot}$, slightly higher than that of SNe \Rmnum{2}P \citep{2017suex.book.....B}. 

Type \Rmnum{2}n SNe are characterized by narrow (tens to hundreds of km\,s$^{-1}$) and/or moderately wide ($\sim 1000$\,km\,s$^{-1}$) hydrogen emission lines in their optical spectra \citep{1990MNRAS.244..269S}. These spectral line features have been attributed to interaction of the ejecta with circumstellar material \citep[CSM;][]{2002MNRAS.330..473C,2014Natur.509..471G}. The brightness rise of SNe \Rmnum{2}n is relatively slow and the time to reach peak brightness is usually more than 20\,days, and SNe \Rmnum{2}n have a mean $\rm{M_{B}\approx -18.7}$ at maximum light \citep{2012ApJ...744...10K}. The light curves of SNe \Rmnum{2}n are diverse; some SN \Rmnum{2}n light curves show long and slow decay, while others have a more rapid linear decline, similar to SNe \Rmnum{2}L. The rapid decline rates observed in some SNe IIn range from 0.03 to 0.07\,mag\,d$^{-1}$, comparable to the steep decline rates seen during the plateau-to-tail transition in SNe IIP.

SNe \Rmnum{2}b are SNe that change from spectral type \Rmnum{2} to spectral type \Rmnum{1}b \citep{1988AJ.....96.1941F}; they are dominated by H lines in early-phase spectra, but the H lines become weak or even disappear at later phases, replaced by P-Cygni profiles of He \Rmnum{1} lines. The nebular spectra are very similar to those of SNe \Rmnum{1}b/c (and SNe \Rmnum{2}P, except for the absence of H lines). The progenitor systems of SNe \Rmnum{2}b may be a single WN-type star \citep[a WR star with highly ionized nitrogen in its spectra;][]{2008MNRAS.391L...5C}, or a binary system in which the progenitor is almost completely stripped of its hydrogen by transfer of matter caused by the wind or companion star. The photosphere thus rapidly recedes to the helium-rich layer below through the small mass of hydrogen-rich ejecta. From then on, SNe \Rmnum{2}b look similar to SNe \Rmnum{1}b.

Some Type II supernovae show narrow emission lines similar to those of Type IIn supernovae in their early stages. However, unlike Type IIn, their narrow emission lines last for a short period. After these narrow emission lines disappear, they are replaced by typical P-Cygni profiles, or the typical P-Cygni profiles appear after a few days of almost featureless continuum spectrum. \cite{2014Natur.509..471G} referred to this rapid disappearance of emission lines as ``flash spectra". Exemplars of this flash spectra type include SN 2013cu \citep{2014Natur.509..471G}, SN 2013fs \citep{2017NatPh..13..510Y}, SN 2015bf \citep{2021MNRAS.505.4890L}, SN 2016bkv \citep{2018ApJ...861...63H, 2018ApJ...859...78N}, SN 2018zd \citep{2020MNRAS.498...84Z}, SN 2023ixf \citep{2023ApJ...956L...5B,Hiramatsu2023ApJ...955L...8H,2023ApJ...956...46S,zhang2023circumstellar,2024Natur.627..759Z}, SN 2024ggi\citep{2024arXiv240607806Z} and so on. 

After the flash emission lines disappear in some SNe \Rmnum{2}, their spectral and photometric evolution is strikingly similar to SNe \Rmnum{2}L. For instance, SN 2015bf \citep{2021MNRAS.505.4890L} exhibits a prominent and broad P-Cygni emission profile with relatively shallow absorption, accompanied by light curves showing an approximately linear decline. According to \cite{2017Natur.551..210A}, the difference between SNe \Rmnum{2}L and SNe \Rmnum{2}P seems to originate more from the cooling properties of the ejecta. However, the influence of CSM may make some SNe \Rmnum{2}P appear like SNe \Rmnum{2}L. Some SNe \Rmnum{2}L have a higher luminosity during the linear decline phase than the plateau phase of SNe \Rmnum{2}P, but their late-time luminosity is consistent with SNe \Rmnum{2}P, such as SN 2018zd, which has a high peak luminosity and a long-term linear decline after the peak, but its late-time luminosity is almost identical to SN 1999em. \cite{2020MNRAS.498...84Z} proposed that the higher luminosity of SN 2018zd during the peak and linear decline phases compared to the plateau of SN 1999em is due to CSM interaction, and the early flash features are produced by the recombination of CSM ionized by high-energy photons.

Research conducted by \cite{2010ApJ...717L..62Y} indicates that certain SNe \Rmnum{2}P may experience a superwind phase prior to their core collapse. \cite{2017ApJ...838...28M} have argued for such super-wind phase and proposed that the interaction between the ejecta and wind resulting from it may explain the differences between SNe \Rmnum{2}L and SNe \Rmnum{2}P light curve morphologies. \cite{2017A&A...605A..83D} pointed out that the phenomena where interaction features briefly appear in the discovery spectra and then fade away within hours or days are actually attributed to the atmosphere or wind closely adjacent to the stellar surface (commonly referred to as the circumstellar envelope), rather than a standard CSM.

Statistical analyses conducted by \cite{2016ApJ...818....3K} and \cite{2021ApJ...912...46B} of spectra obtained within 5 days and 2 days post-explosion (respectively) revealed that over $18\%$ and $30\%$ of these early-time spectra exhibited flash features. Attempts were also made to demonstrate that supernovae (SNe) with flash features are more luminous, bluer, or rise more slowly than those without such features, but statistical analyses do not support this conclusion \citep{2021ApJ...912...46B,2023ApJ...952..119B}. \cite{2023ApJ...952..119B} measured the persistence duration of flash-ionization emission and find that most SNe show flash features for $\approx$ 5 days. However, those rarer events, with persistence timescales $\le$ 10 days, are brighter and rise longer, suggesting these may be intermediate between regular SNe II and strongly interacting SNe \Rmnum{2}n. An accurate estimate of the fraction and confirmation of the above tendency require a sample with observations in the first few hours after explosion when the shock wave breaks out of the stellar envelope\citep{2024Natur.627..754L}.

The origin of flash emission lines is attributed to the CSM surrounding the progenitor star. This CSM undergoes ionization by high-energy photons or shockwaves, followed by a rapid cooling and recombination process, resulting in the generation of these flash emission lines. As demonstrated by \cite{2016ApJ...818....3K} and \cite{2021ApJ...912...46B}, flash-spectra emission lines are prevalent in the earliest stages of SN explosions. These emission lines offer valuable insights into the composition and structure of the CSM, thereby serving as a valuable tool for understanding the late-stage evolutionary activity of massive stars, mass-loss processes, CSM formation, and the interaction between ejecta and the CSM.

The initial two spectra of SN 2021dbg exhibit prominent flash emission lines that completely disappeared within 5 days, suggesting the possible presence of a certain amount of CSM near its progenitor, but not in the large quantities seen in Type \Rmnum{2}n supernova. SN 2021dbg also stands out among SNe \Rmnum{2} owing to its exceptional luminosity. Its light curve and spectral features exhibit a blend of SNe \Rmnum{2}P and SNe \Rmnum{2}L characteristics, suggesting that it may occupy a transitional zone between these two SN types. Such SNe, positioned at the boundary of typical classifications, offer unique opportunities to unravel the physical mechanisms underlying the formation of different SN types and the final evolutionary processes of progenitors with varying main-sequence masses.

In this paper, optical photometry and spectra of SN 2021dbg are analyzed comprehensively. Section 2 shows our data sources and processing methods. In Section 3, we give a comprehensive analysis of the light curves and optical spectra, including the comparison with other well-studied SNe \Rmnum{2}. Parameters of SN 2021dbg's progenitor are discussed in Section 4, such as the mass of the progenitor, $^{56}$Ni mass, mass loss, and material distribution structure. In Section 5, the nature of SN 2021dbg and some conclusions are briefly summarized.

\section{Observations} 
\subsection{Discovery}
SN 2021dbg (ATLAS21gyf) was reported by the Asteroid Terrestrial-impact Last Alert System (ATLAS; \citealp{2018PASP..130f4505T, 2018ApJ...867..105T, 2018AJ....156..241H})  and was first detected at J2000 coordinates $\alpha = 09^{\rm h}24^{\rm m}26.801^{\rm s}$, $\delta = -06^\circ34'53.09''$, on 2021-02-15 10:14:52.800 (UTC dates are used throughout this paper). The detected brightness measured with the ATLAS-01 {\it cyan (c)} filter was 18.13 (AB magnitude). ATLAS also conducted subsequent observations, acquiring photometric data in the {\it c} and {\it o} bands. The extended Public European Southern Observatory (ESO) Spectroscopic Survey of Transient Objects (ePESSTO) team \citep{2021TNSCR.489....1H} observed the initial spectrum using EFOSC2 on the ESO-NTT on 2021-02-16 at 04:05:44. This observation confirmed SN 2021dbg as an SN~II. In response, we promptly utilized the 2.4\,m Li-Jiang telescope (hereafter referred to as LJT; \citealp{2015RAA....15..918F}) at the Li-Jiang Observatory of Yunnan Observatories equipped with the YFOSC (Yunnan Faint Object Spectrograph and Camera; \citealp{2019RAA....19..149W}) for continuous photometric and spectral observations.

\subsection{Photometry}
All images obtained by LJT were processed according to IRAF standard procedures, including template subtraction, background subtraction, flat-fielding, and cosmic ray removal. Instrumental magnitudes were obtained by using the point-spread-function (PSF) fitting method \citep{1987PASP...99..191S}. The instrumental magnitudes were then converted into standard Johnson $BV$ \citep{1966CoLPL...4...99J} and standard Sloan $gri$ (AB magnitude). 

\begin{figure*}
\centering \includegraphics[width=1.0\linewidth]{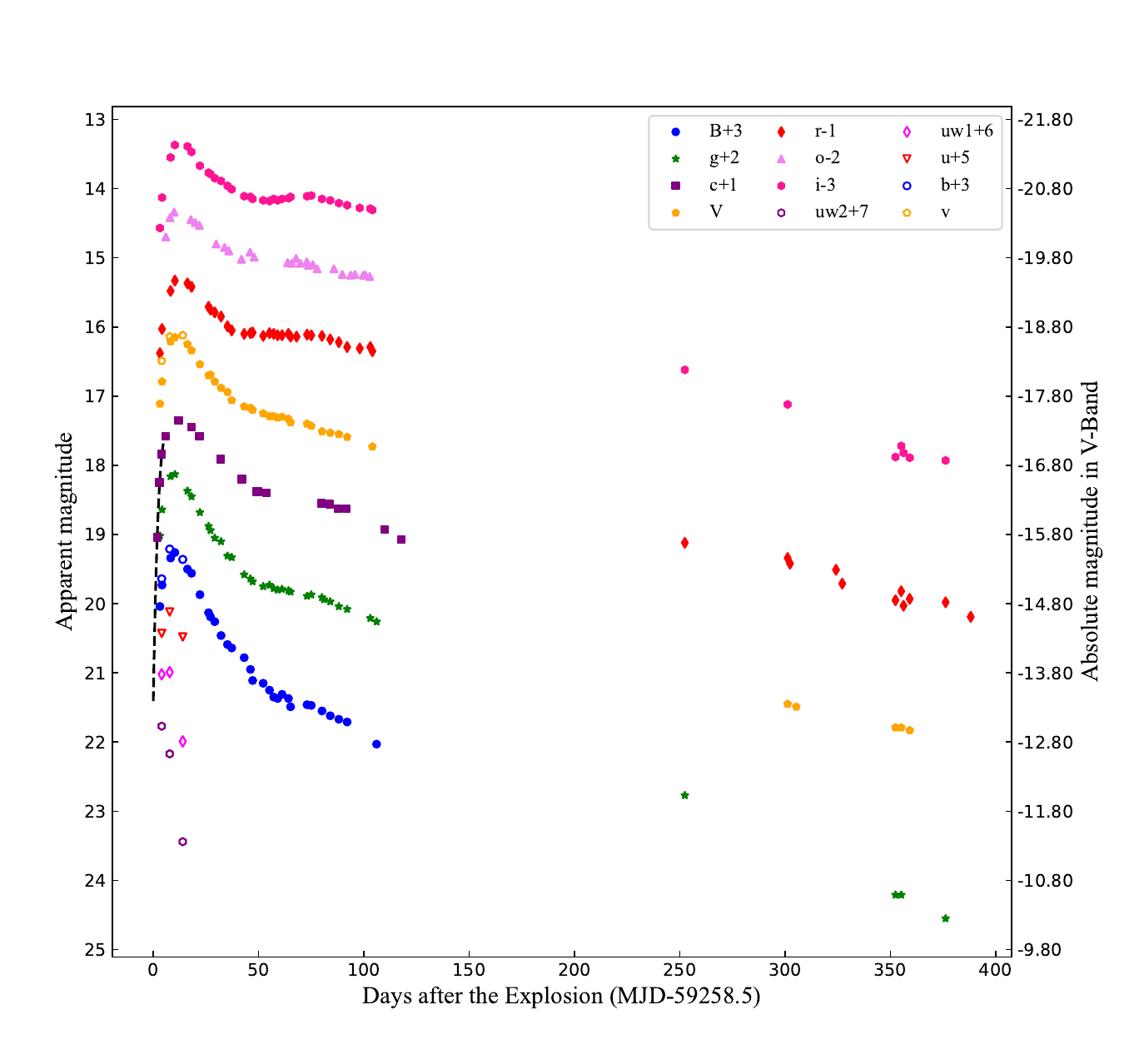}
\caption{Optical light curves of SN 2021dbg, where $B$, $V$, $g$, $r$, and $i$ were obtained with the 2.4\,m LJT; $c$ and $o$ were obtained with ATLAS-MLO+ATLAS-01; and $uw2$, $uw1$, $u$, $b$, and $v$ were obtained with {\it Swift}-UVOT. All measurements are shifted vertically for clarity. The black dashed curve shows the fit to the $c$ band at early times to estimate the explosion time. }
\label{fig:1}
\end{figure*}

SN 2021dbg was observed by the Ultraviolet/Optical Telescope (UVOT) of the {\it Swift} Observatory in the $uvw2$, $uvw1$, $um2$, $u$, $b$, and $v$ bands, but only three images near the luminosity peaks were obtained in each band. Figure \ref{fig:1} displays light curves in all bands we obtained. Table \ref{tab:A1} presents our all photometric data obtained by LJT, ATLAS and {\it Swift}. Figure \ref{fig:A1} shows the finder field of SN 2021dbg and local reference stars, whose information is listed in Table \ref{tab:A2}.

Because of the first detection, we used a simple expanding fireball model, $F(t)=A\times(t-t_{0})^{2}+B$, where $A$ and $B$ are fitting coefficients, to fit the early detections of SN 2021dbg in the {\it cyan} band observed by ATLAS to estimate the explosion epoch. The fitting result is shown with the black dashed curve in Figure \ref{fig:1}. We adopt MJD = $59258.5 \pm 0.5$ as our estimated explosion time, which is 1.9 days prior to the first detection.

\subsection{Spectroscopy}
\begin{figure*}
\plotone{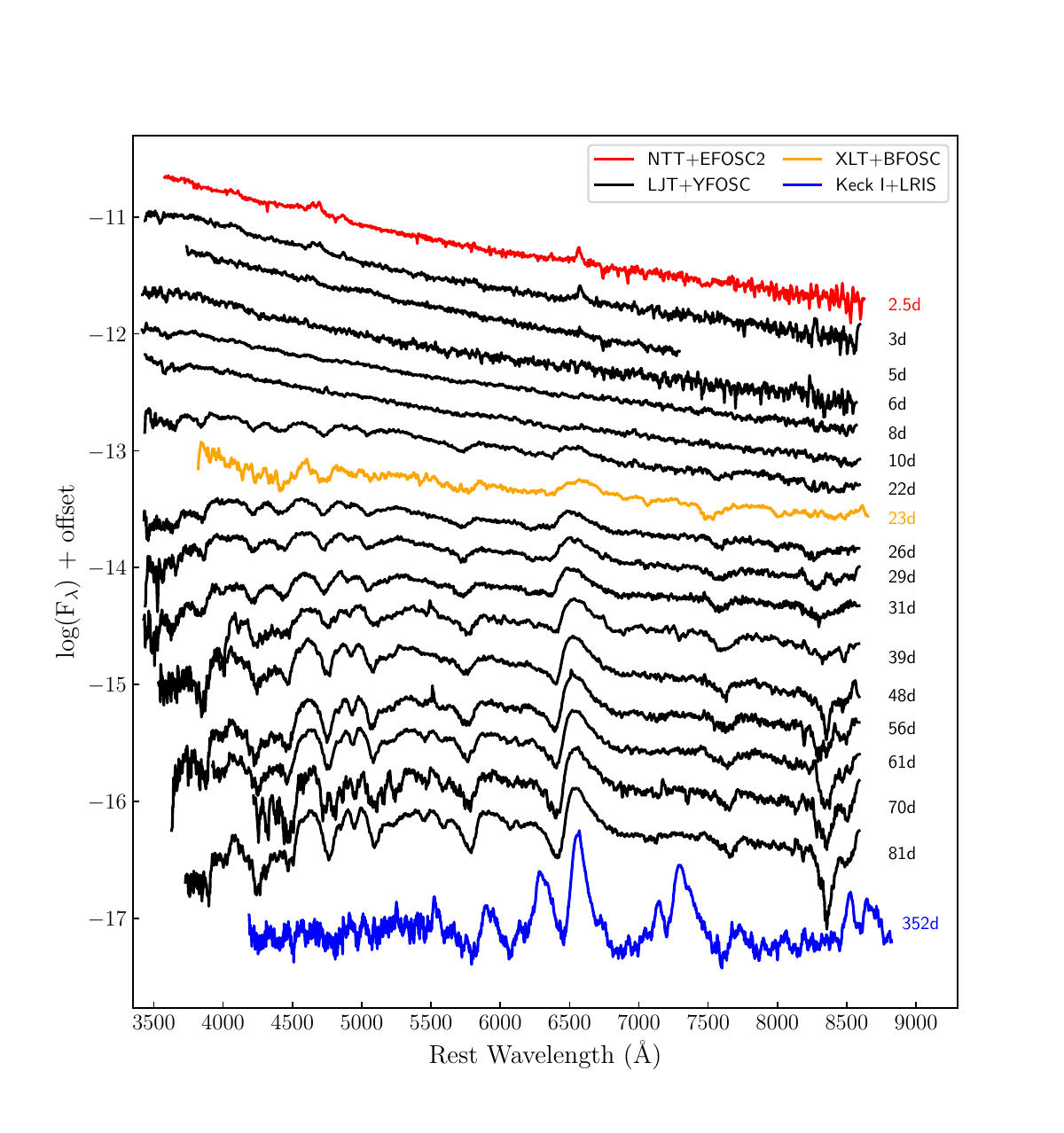}
\caption{Spectra of SN 2021dbg obtained with LJT+YFOSC (black), NTT+EFOSC2 (red), XLT+BFOSC (orange), and Keck~I+LRIS (blue); The time on the right of each spectra is relative to the explosion time we estimated.  All spectra were corrected for redshift and were vertically shifted for clarity.}
\label{fig:2}
\end{figure*}

LJT captured a total of 15 spectra spanning a period of $\sim 3$ to 81 days post-explosion. The wavelengths and fluxes of these spectra were calibrated according to IRAF standard procedures, telluric absorption was \textbf{removed} by comparison with standard-star spectra, and atmospheric extinction correction was carried out according to the atmospheric extinction information of local stations. Several spectra with large flux-calibration deviations were corrected with multiband photometry. In addition, we used the XingLong 2.16\,m telescope (+ BFOSC; at Xing-Long Observatory of the National Astronomical Observatories of China; hereafter referred to as XLT) to take a spectrum $\sim 23$ days after the explosion. The Low-Resolution Imaging Spectrometer \citep[LRIS;][]{1995PASP..107..375O} on the Keck~I 10\,m telescope was used to take a spectrum of the nebular phase $\sim 352$\,days after the explosion; use of an atmospheric dispersion corrector (ADC) precluded differential slit losses \citep{1982PASP...94..715F}. Table \ref{tab:A3} lists the journal of 
spectral observations. Figure \ref{fig:2} displays the spectra observed by the LJT+YFOSC, NTT+EFOSC2, XLT+BFOSC, and Keck~I+LRIS. From Figure \ref{fig:2}, we can see that the first three spectra exhibit weak flash features, followed by a disappearance of the flash features in the next few spectra, forming a blue featureless phase. Then, low-contrast H lines emerge, and it is until about 30 days after the explosion that clear photospheric features (P-Cygni) appear.

\section{Analysis and Results} 
No Cepheid variable distance or rotation velocity have been reported for the host galaxy, precluding a direct estimation of the distance to SN 2021dbg. Instead, we rely on the redshift value of $z = 0.02$ \citep{2021TNSCR.489....1H,2021TNSCR.501....1Z} and the cosmological parameters from the Nine-year Wilkinson Microwave Anisotropy Probe (WMAP9; \citealt{2013ApJS..208...19H}) to calculate a distance of 90\,Mpc. Limited by our low spectral resolution, we cannot estimate the host-galaxy extinction from the absorption of Na~\Rmnum{1}~D $\lambda\lambda  5890$, 5896\,\AA\ lines. For the Milky Way extinction, we use the extinction data provided by \cite{2011ApJ...737..103S}  in NASA's Extragalactic Database\footnote{\url{https://ned.ipac.caltech.edu/}}, $E(B-V)_{\rm MW} = 0.032$\,mag, $A_V = 0.087$\,mag. The light curve and spectrum of SN 2021dbg in this paper are only corrected for Galactic extinction.

\subsection{Photometric Analysis}
\subsubsection{Evolution and comparison of light curves}

SN 2021dbg attains its peak luminosity nearly concurrently in all optical bands. The light curve for the $B$ and $g$ bands exhibits a linear decline from the peak until 50 days post-explosion, subsequently transitioning into a prolonged plateau phase of slower decline. Meanwhile, the $c$, $V$, $r$, $o$, and $i$ bands enter this plateau phase earlier, at 43 days. These plateaus persist until $\sim 110$ days after the explosion. Notably, during the plateau phase, there is a subtle brightening in the $r$, $o$, and $i$ bands, with increases of $\sim 0.2$\,mag in the $r$ and $o$ bands and a more pronounced brightening of $\sim 0.7$\,mag in the $i$ band, resulting in a distinct bump in the light curve.

SN 2021dbg stands out as a relatively bright object among Type \Rmnum{2} SNe. Its absolute peak magnitudes exceed $-18.4$ in all optical bands and $-19.8$ in all UV bands. Table \ref{tab:1} provides peak information and the decline rate from peak to plateau on the optical light curves. The optical and UV bands of SN 2021dbg are brighter than those of typical well-studied Type \Rmnum{2}P SNe, fall within the intermediate level for Type \Rmnum{2}L SNe, and are slightly fainter than the median brightness for Type \Rmnum{2}n SNe. Previously, the optical absolute magnitude range for Type \Rmnum{2}P SNe  was reported as $-18 \le M_{\rm op} \le -15$\,mag, while for Type \Rmnum{2}n SNe, it was $-20 \le M_{\rm op} \le -18$\,mag. In UV bands, Type \Rmnum{2}P SNe had an absolute magnitude of $M_{\rm UV} \approx -18$\,mag, and Type \Rmnum{2}n SNe had $M_{\rm UV} \approx -20$\,mag \citep{2014ApJ...787..157P}. \cite{2002AJ....123..745R} conducted a statistical analysis of the brightness in optical bands for Type \Rmnum{2}P and Type \Rmnum{2}L SNe, finding average peaks of $M_B = -17.0 \pm 1.1$\,mag for Type \Rmnum{2}P and $M_B = -18 \pm 0.9$\,mag for Type \Rmnum{2}L.

The light curves of SN 2021dbg during the first 50 days resemble those of typical SNe~\Rmnum{2}L, characterized by high brightness and a linear decline after the peak. However, after 50 days, the light curves exhibit a prominent plateau, a hallmark feature of SNe~\Rmnum{2}P. If Type \Rmnum{2}P and Type \Rmnum{2}L SNe are considered two ends of a continuous family of Type \Rmnum{2} SNe, SN 2021dbg occupies the brighter end of the transitional zone.

\begin{figure*}
\gridline{
  \fig{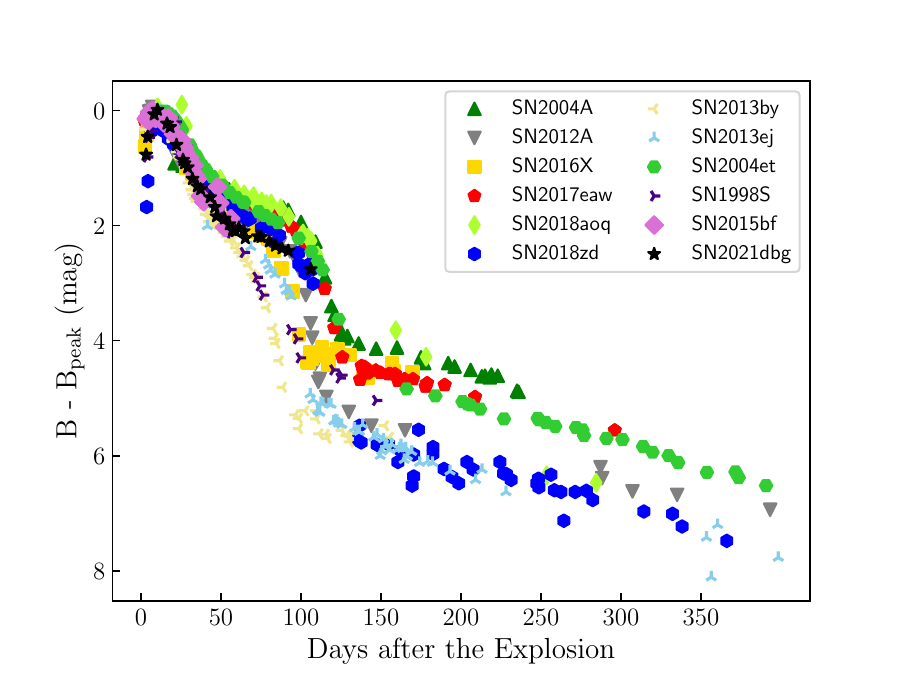}{0.48\textwidth}{(a)}
  \fig{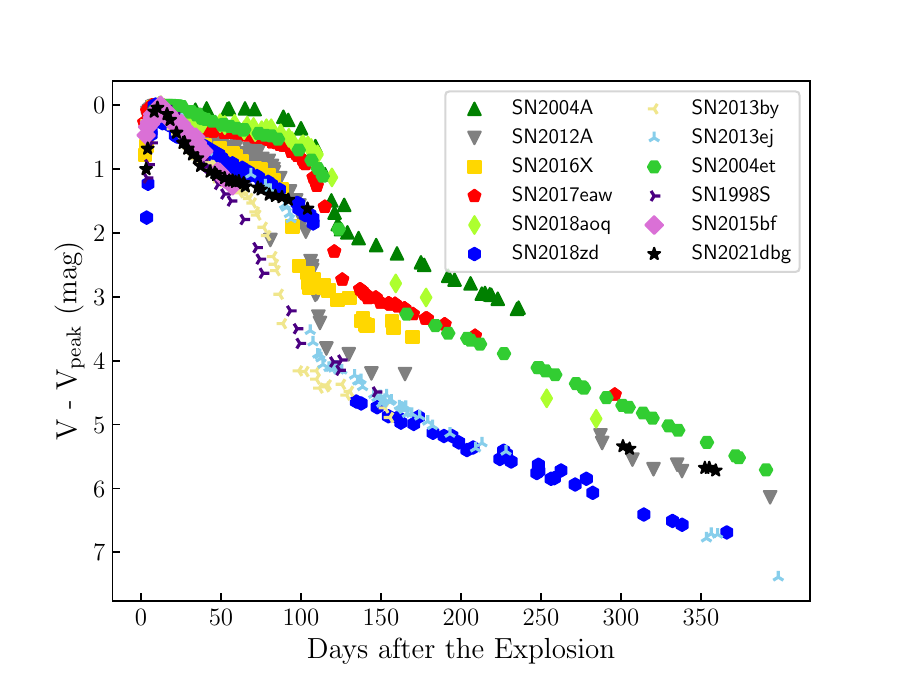}{0.48\textwidth}{(b)}
         }
\gridline{
  \fig{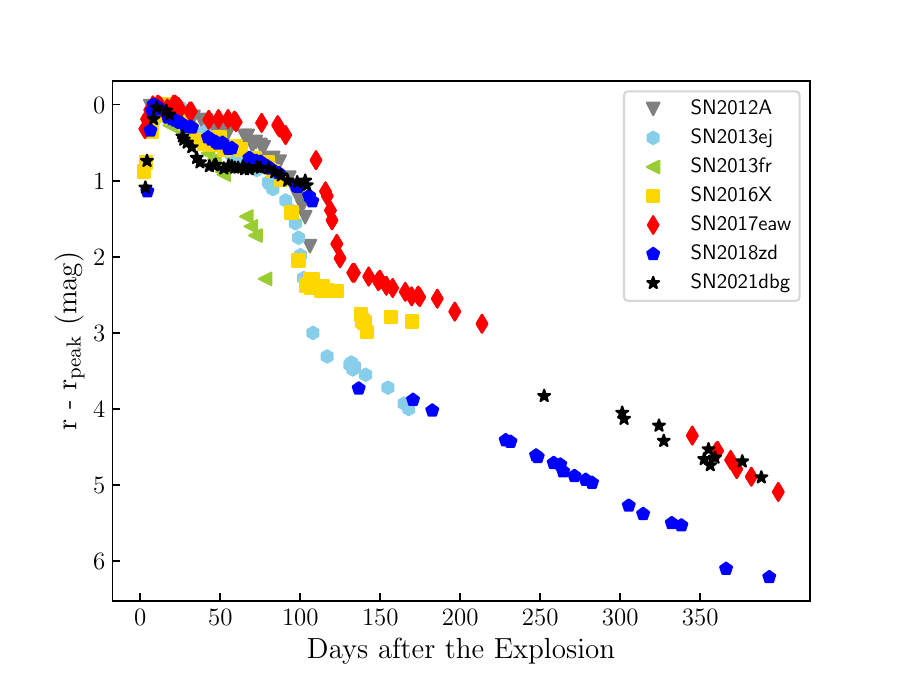}{0.48\textwidth}{(c)}
  \fig{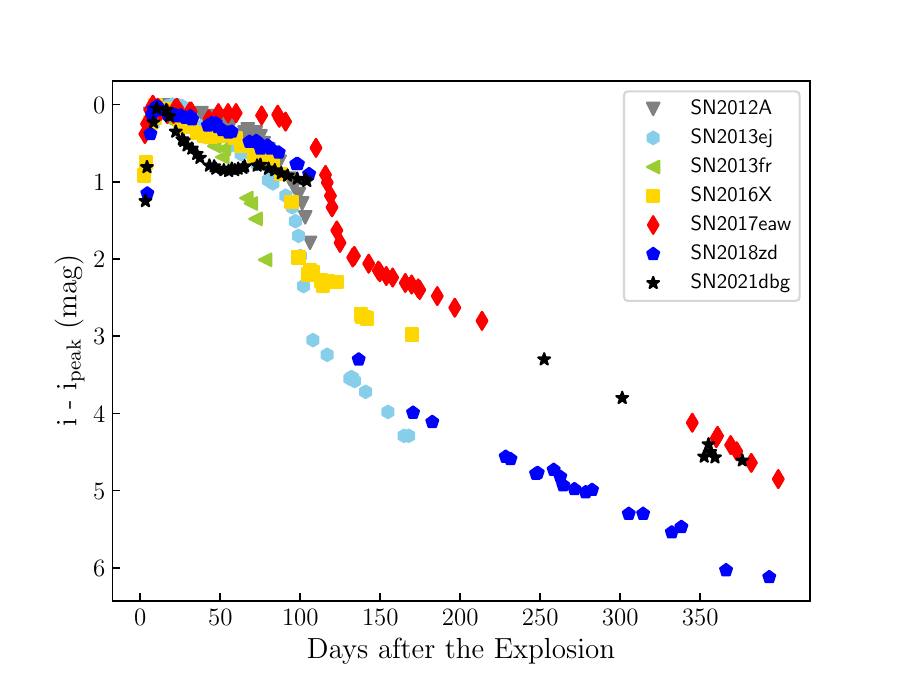}{0.48\textwidth}{(d)}
         }
\caption{Panels (a), (b), (c), and (d) present $B$, $V$, $r$, and $i$ light curves of SN 2021dbg and other SNe \Rmnum{2} respectively, which are matched at peak brightness.}
\label{fig:3}
\end{figure*}

To better analyze the evolution characteristics of the SN 2021dbg light curves, we compared them with those of some other well-studied SNe \Rmnum{2} after matching the peak values; see Figure \ref{fig:3}. 

In the $B$ band, the decline rate from peak luminosity to 50 days post-explosion is 0.046\,mag\,d$^{-1}$ for SN 2021dbg. This rate aligns with that observed in SN 2012A \citep{2013MNRAS.434.1636T}, SN 2016X \citep{2018MNRAS.475.3959H}, and SN 2015bf \citep{2021MNRAS.505.4890L}, placing it firmly within the mid-range of SNe \Rmnum{2} samples in Figure \ref{fig:3}. However, several SNe, including SN 2004A \citep{2008BASI...36...79G}, SN 2004et \citep{2006MNRAS.372.1315S}, SN 2017eaw \citep{2019ApJ...875..136V}, SN 2018zd, and SN 2018aoq \citep{2021AstL...47..291T}, exhibit a slower decline and transition into a plateau phase sooner than SN 2021dbg. Conversely, SN 1998S \citep{2000MNRAS.318.1093F}, SN 2013by \citep{2015MNRAS.448.2608V}, and SN 2013ej \citep{2016ApJ...822....6D} decline more rapidly but lack a discernible plateau phase.

In the $V$ band, the light curve of SN 2021dbg decreases by 1.19\,mag from peak to 50 days, with a decline rate of 0.029\,mag\,d$^{-1}$, which is the fastest among all SN \Rmnum{2}P samples; only SN 2013by is comparable, but SN 2013by does not have a significant plateau after that. \cite{2015MNRAS.448.2608V} classified SN 2013by as a young SN \Rmnum{2}L/\Rmnum{2}n based on early optical and NIR observations. SN 2013ej, SN 2015bf, SN 2016X, and SN 2018zd also do not have a significant plateau, declining linearly nearly from the peak until $\sim 90$--110 days, after which, like all SNe \Rmnum{2}P, they rapidly decline to the tail phase. Although an obvious decline before the transition to the $^{56}$Ni decay tail phase is the feature of typical SNe \Rmnum{2}P, \cite{2014ApJ...786...67A} and \cite{2015MNRAS.448.2608V} proposed that all SNe \Rmnum{2}L monitored long enough (more than $\sim 80$ days since discovery) demonstrated this decline. 

In both the $r$ and $i$ bands, from the peak to the plateau period, the brightness declines 0.82\,mag, with a decline rate of 0.022\,mag\,d$^{-1}$. Compared to typical SNe \Rmnum{2}P, such as SN 2012A, SN 2016X, and SN 2017eaw, the SN 2021dbg peak has a larger peak-to-plateau drop, and takes longer to decline from peak to plateau. This behavior can be explained by interactions between ejecta and the CSM. In the plateau and tail phases, SN 2021dbg is consistent with typical SNe \Rmnum{2}P. 

\begin{figure*}
\gridline{
  \fig{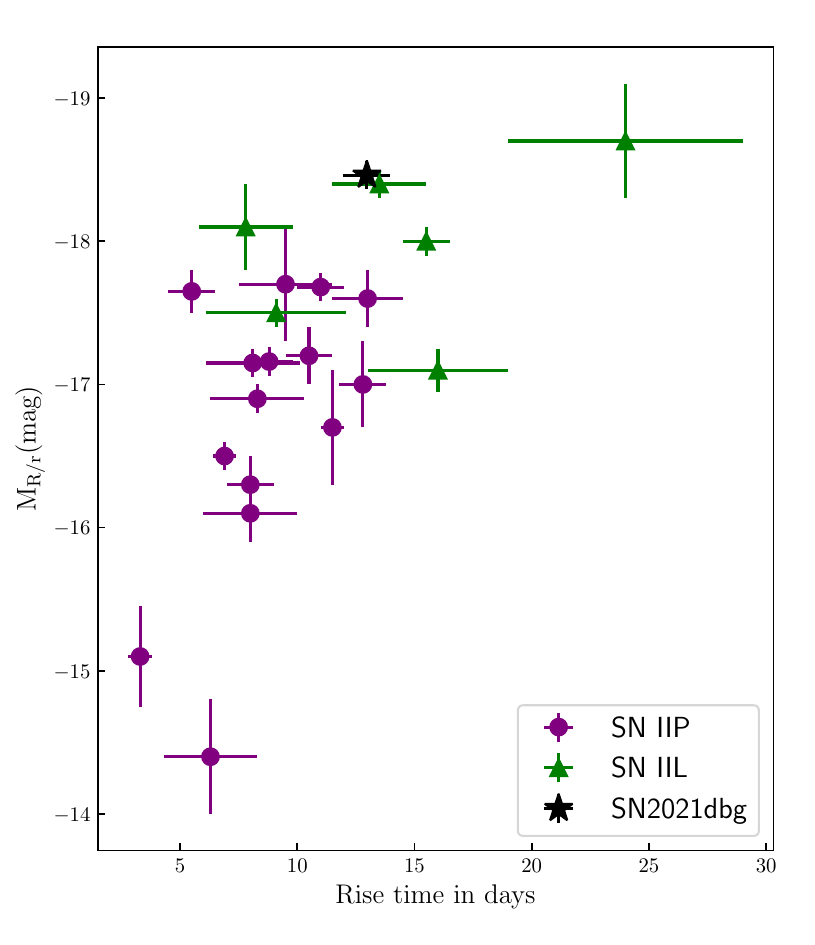}{0.315\textwidth}{(a)}
  \fig{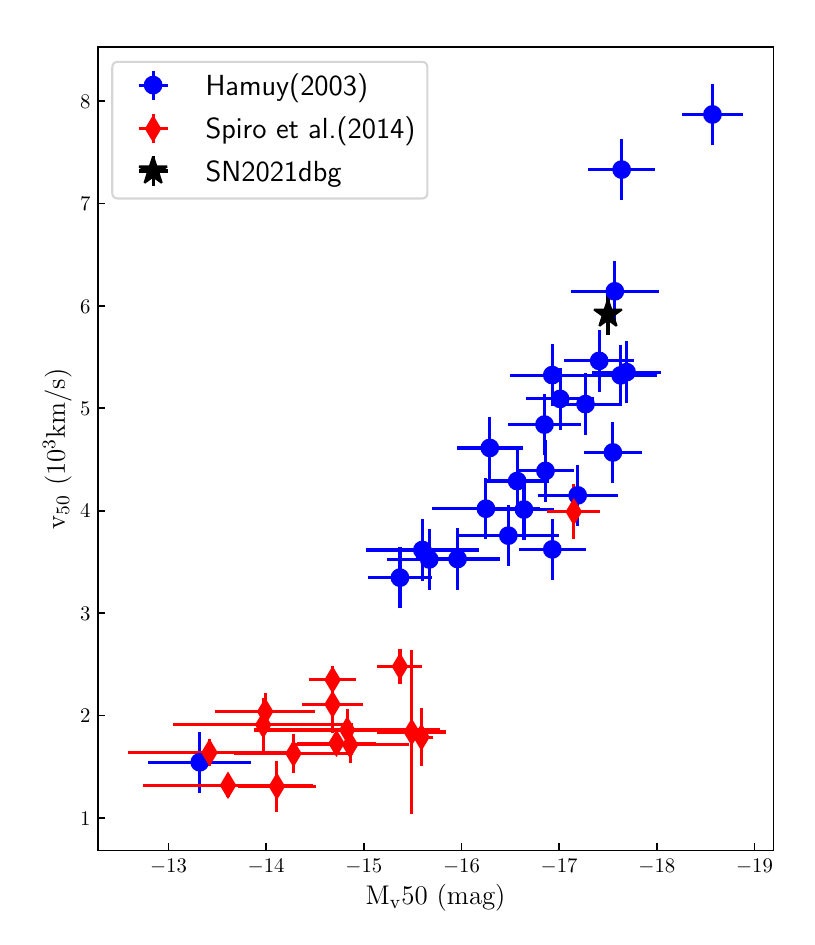}{0.315\textwidth}{(b)}
  \fig{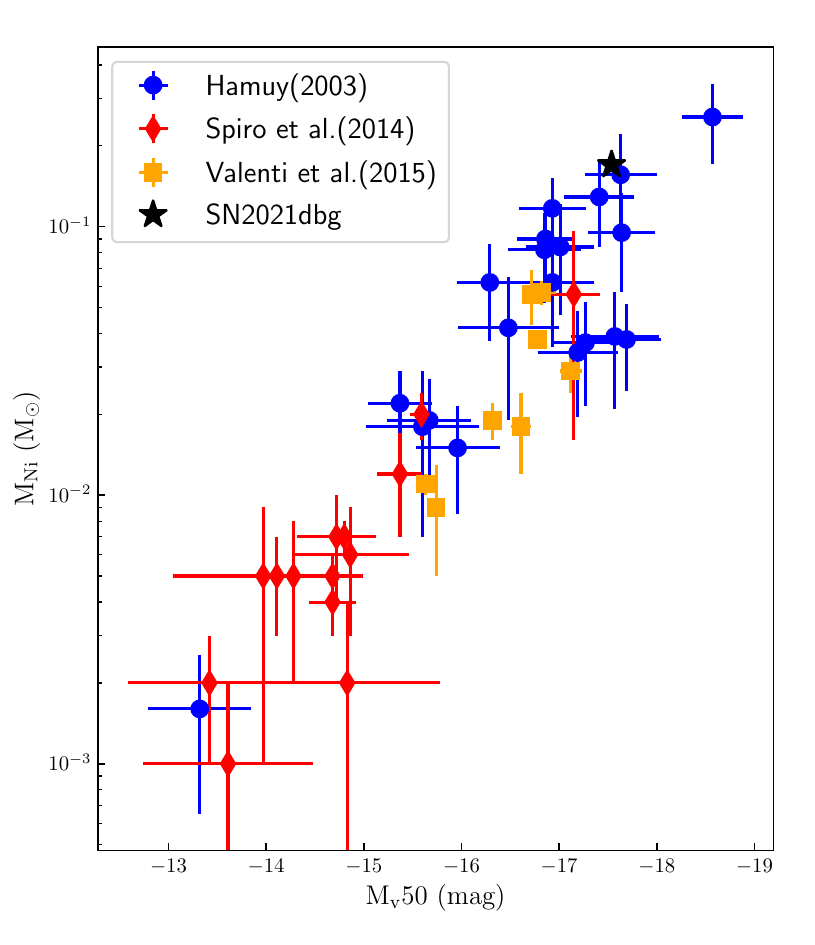}{0.315\textwidth}{(c)}
         }
\caption{The position of SN 2021dbg within the SN \Rmnum{2} population in terms of various photometric indicators. {\it Panel (a):} $R/r$-band absolute magnitudes at the end-of-rise epoch and rise time; The samples of SNe \Rmnum{2}P and SNe \Rmnum{2}L are from \citep {2015A&A...582A...3G}. {\it Panel (b):} Velocity of  Fe~\Rmnum{2} $\lambda$5169\,\AA\ measured at 50 days ($v_{50}$) and the $V$-band absolute
magnitude measured at 50 days ($M_{V}50$
). The blue pots represent the SNe \Rmnum{2}p samples collected from \cite{2003ApJ...582..905H}, the red pots represent the low luminosity SNe \Rmnum{2}p samples collected from \cite{2014MNRAS.439.2873S}. {\it Panel (c):} $V$-band absolute
magnitude measured at 50 days ($M_{V}50$), and the $^{56}$Ni mass. The orange pots represent the SNe \Rmnum{2}P and SNe \Rmnum{2}L samples collected from \cite{2015MNRAS.448.2608V}.}
\label{fig:4}
\end{figure*}

Figure \ref{fig:4} shows the position of SN 2021dbg within the SN \Rmnum{2} population in terms of various photometric indicators.
In the left panel, we can see that the brightness of SN 2021dbg is brighter than the SNe \Rmnum{2}P samples in the panel, placing it among the SNe \Rmnum{2}L samples. In the middle and right panels, we can observe that SN 2021dbg is positioned towards the top of the SNe \Rmnum{2}P samples provided by \cite{2003ApJ...582..905H}. These indicators of SN 2021dbg suggest that it possesses the high brightness of SNe \Rmnum{2}L, yet does not completely deviate from the range of SNe \Rmnum{2}P, implying that SN 2021dbg may be a transitional supernova event between SNe IIP and SNe \Rmnum{2}L.

\begin{deluxetable*}{cccccccc}
\tablenum{1}
\tablecaption{Information about the peaks of light curves}
\label{tab:1}
\tablewidth{0pt}
\tablehead{
\nocolhead{Common} & \colhead{$B$} &\colhead{$V$} & \colhead{$g$} & \colhead{$r$} & \colhead{$i$} & \colhead{$c$} & \colhead{$o$}
}
\startdata
Peak	Time(d)	&	10.93	&	11.69	&	10.82	&	12.97	&	12.70	&	11.86	&	12.22\\	
Peak (mag)	&	-18.58	&	-18.66	&	-18.69	&	-18.46	&	-18.40	&	-18.48	&	-18.45	\\	
M50 (mag)	&	-16.77	&	-17.54	&	-17.08	&	-17.66	&	-17.57	&	-17.33	&	-17.67	\\	
Decline	rate (mag/d)	&	0.046	&	0.033	&	0.041	&	0.027	&	0.026	&	0.031	&	0.023	\\
\enddata
\tablecomments{M50 is the absolute magnitude in 50 days; the decline rate is calculated from peak to 50 days.}
\end{deluxetable*}

\subsubsection{Evolution and comparison of color}
\begin{figure*}
\gridline{
  \fig{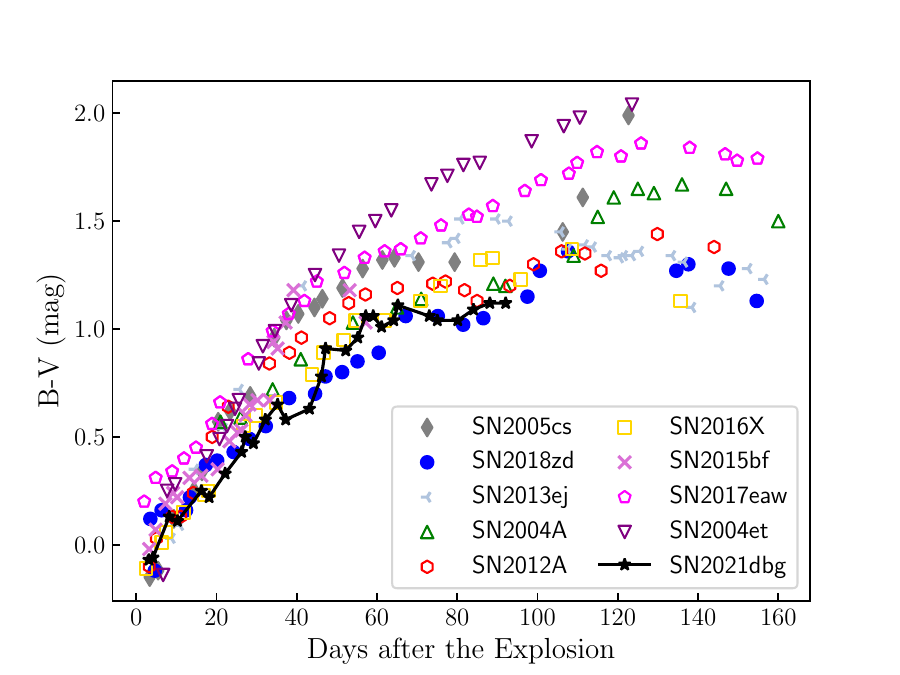}{0.5\textwidth}{(a)}
  \fig{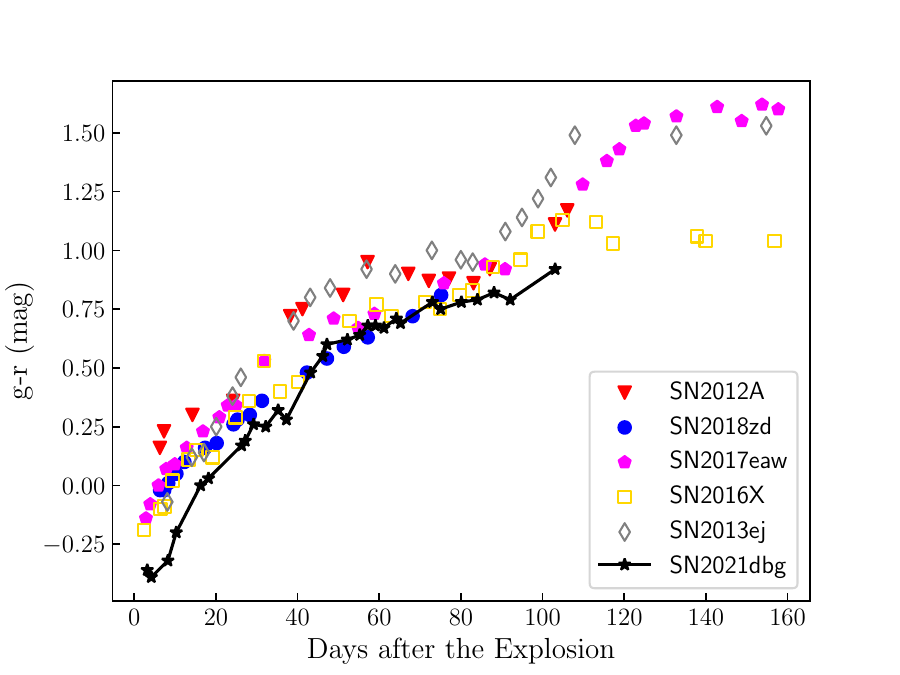}{0.5\textwidth}{(b)}
         }

\caption{SN 2021dbg's color comparison with that of other SNe \Rmnum{2}. Panel (a) and (b) present the B-V color and g-r color respectively.}
\label{fig:5}
\end{figure*}

Since SN 2021dbg exhibits features of both SNe \Rmnum{2}P and SNe \Rmnum{2}L in its light curves and spectra, we have selected some well-studied and typical SNe \Rmnum{2}P and SNe \Rmnum{2}L as comparison samples. Some of them show flash emission features and CSM interaction. Figure \ref{fig:5} shows SN 2021dbg's color evolution and color comparison with these SNe \Rmnum{2} samples.

The color evolution trend of SN 2021dbg is consistent with that of typical SNe \Rmnum{2}P, but bluer, which may be the result of CSM interaction. The color curve of SN 2021dbg is very similar to that of SN 2016X and SN 2018zd, which means that they may have comparable CSM interaction. The $B-V$ color curve of SN 2021dbg is very similar to that of SN 2016X, SN 2018zd, and SN 2018aoq during the first 70 days, but SN 2021dbg and SN 2018zd show a brief bluing feature at 70--80 days and then slowly turn red. Similar to SN 2016X, SN 2013ej, and SN 2012A, the $B-V$ color curve flattens out and slowly turns blue after $\sim 110$ days, while SN 2004A and SN 2017eaw gradually turn blue after $\sim 125$ days, and SN 2018aoq continues to turn red linearly until $\sim 170$ days \citep{2021AstL...47..291T}. The $uvw2-u$ color curve of SN 2018zd exhibits a ``U-turn'' feature at early times, indicating that SN 2018zd had a temperature increase  and the dust may be destroyed by shock waves to reduce extinction \citep{2020MNRAS.498...84Z}.
For the first 40 days, SN 2021dbg's $g-r$ color is bluer than that of SN 2016X and SN 2018zd, and has been almost always at the bluest end of these SNe II samples in Figure \ref{fig:5}.

If such a blue light curve is indeed driven by CSM interaction, it must be exceptionally intense, requiring a significant amount of CSM, similar to the case in SN 2018zd \citep{2020MNRAS.498...84Z}. However, this appears to be in contrast with the flash emission lines that disappeared only around five days after the explosion, and the typical rapid cooling observed when CSM interaction alone drives the early light curve. As discussed in Section \ref{Sect 4.3}, we propose that the prolonged bluer light curve of SN 2021dbg may be associated with the presence of a stratified structure in the progenitor's envelope.

\subsection{Spectral Analysis}
\subsubsection{Evolution and comparison of early-phase spectra}

Early-time spectra, especially those showing flash emission lines 
\citep[e.g.,][]{2014Natur.509..471G}, can help us understand the composition and structure of matter near the progenitor. To explore the relationship between flash emission lines in early spectra and  matter near the progenitor, \cite{2017A&A...605A..83D} constructed a generation model of  spectra and successfully reproduced the flash emission lines. 

\begin{figure*}
\plotone{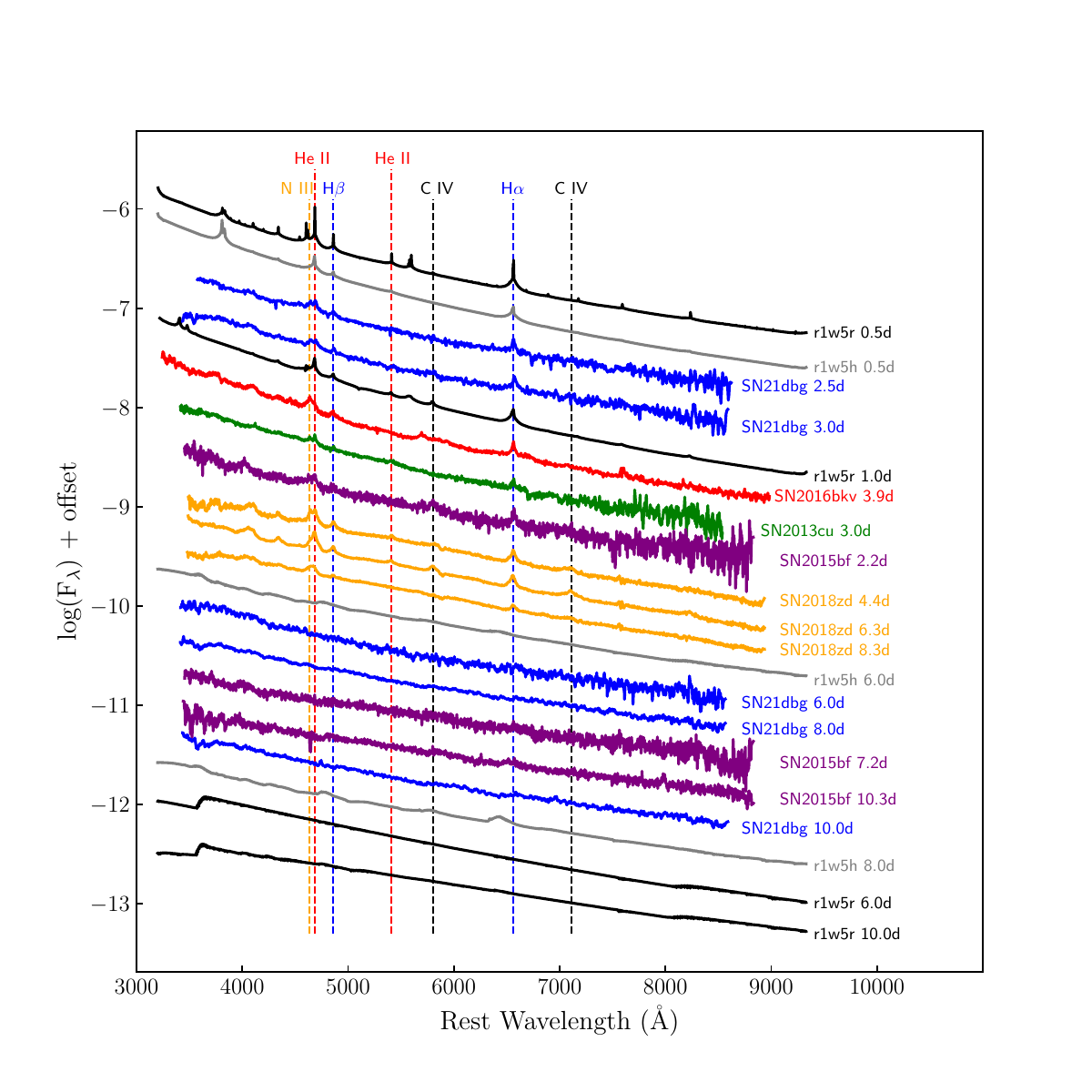}
\caption{Early-time spectra of SN 2021dbg compared with those of other SNe \Rmnum{2} with flash emission lines. Black and gray spectra are r1w5r and r1w5h model spectra\citep{2017A&A...605A..83D}. The time on the right of each spectra is relative to the explosion time. All spectra were corrected for redshift and  vertically shifted for clarity.}
\label{fig:6}
\end{figure*}

Figure \ref{fig:6} shows the early-time spectra of SN 2021dbg and several other SNe \Rmnum{2}, as well as the spectra of the r1w5h and r1w5r models \citep{2017A&A...605A..83D}. The spectra of SN 2021dbg 2.5 and 3 days after the explosion exhibit narrow H$\alpha$, H$\beta$, He~\Rmnum{2} $\lambda 4686$~\AA, N~\Rmnum{3} $\lambda\lambda4334$,  4641~\AA, and weak C~\Rmnum{4} $\lambda\lambda 5801$, 5812\,\AA\ and C~\Rmnum{4} $\lambda 7110$~\AA. These flash emission lines are the same as those of SN 2013cu, SN 2015bf, SN 2016bkv, and SN 2018zd, and they also agree with the emission lines produced by the r1w5r model. They are thought to result from the recombination of stellar-wind material near the progenitor after being ionized by X-rays from shocked ejecta. SN 2013cu showed narrow and strong  H$\alpha$, H$\beta$, He~\Rmnum{1}, He~\Rmnum{2}, C~\Rmnum{3}, C~\Rmnum{4}, and N~\Rmnum{4} emission lines within 15.5\,hr after the explosion \citep{2014Natur.509..471G}, while by 3 days, He~\Rmnum{1}, C~\Rmnum{3}, C~\Rmnum{4}, and N~\Rmnum{4} emission lines disappeared, leaving only weak H$\alpha$, H$\beta$, and He~\Rmnum{2} lines. Therefore, SN 2021dbg may have had strong and narrow flash emission lines as did SN 2013cu, although we did not obtain SN 2021dbg's spectra at sufficiently early times after the explosion. At 6--10 days, the spectra of SN 2021dbg do not exhibit any significant emission lines except for very weak H$\alpha$, consistent with SN 2015bf and also with spectra of the r1w5h and r1w5r models.

In the model by \cite{2017A&A...605A..83D}, w5 represents a mass loss rate of $5\times 10^{-3}M_{\odot} yr^{-1}$, which indicates that the progenitor of SN 2021dbg may have experienced an enhanced mass loss shortly before the explosion, or undergone a superwind phase proposed by \cite{2010ApJ...717L..62Y}. We discuss the mass loss of the progenitor in more detail in Section \ref{Sect 4.4}.

\begin{figure*}
\plotone{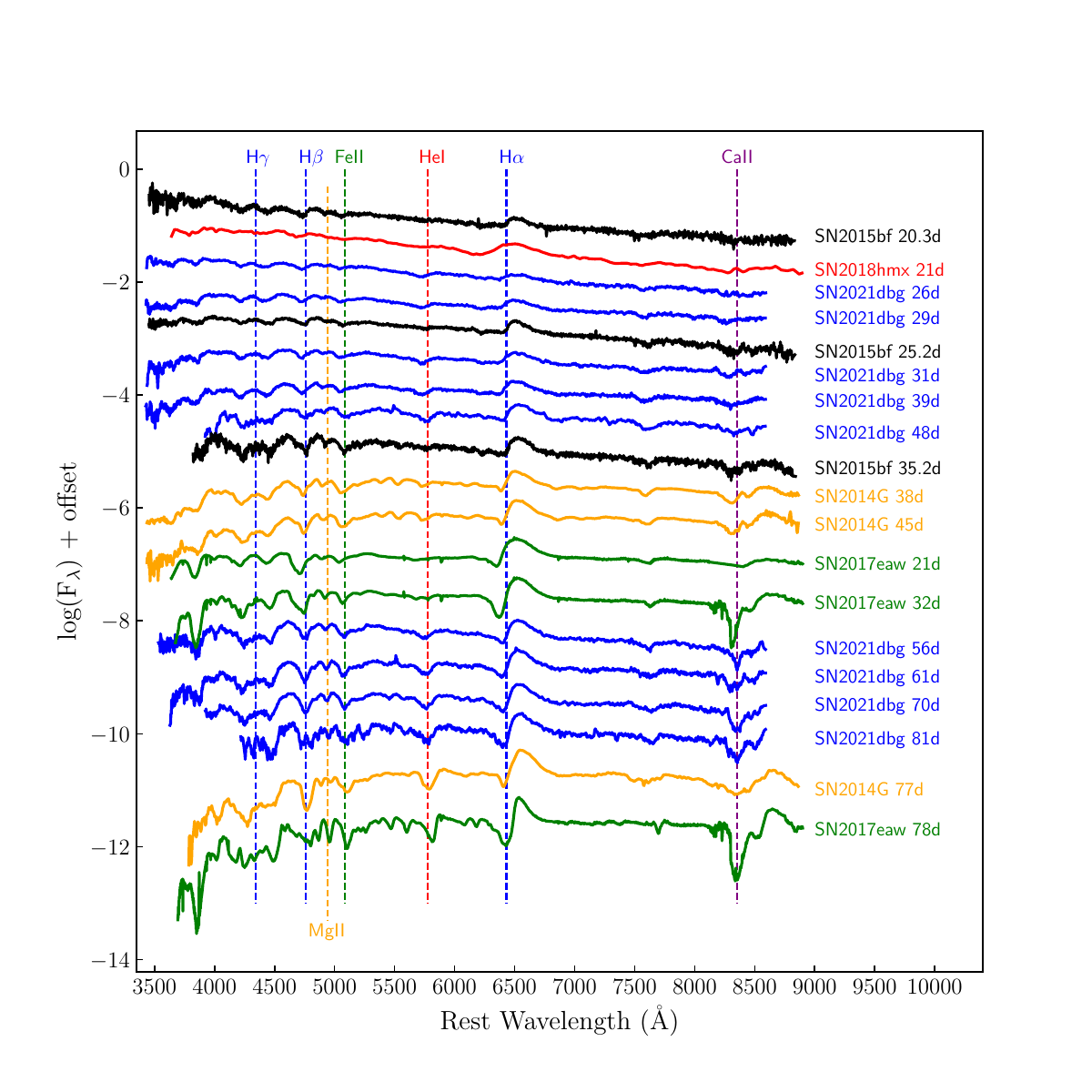}
\caption{Photospheric-phase spectra of SN 2021dbg compared with those of other SNe \Rmnum{2}. The time on the right of each spectra is relative to the explosion time. All spectra were corrected for redshift and  vertically shifted for clarity. The dashed lines indicate the spectral features at certain velocities, where H$\alpha$ is located at -6000 \,km\,s$^{-1}$, H$\beta$ at -6000 \,km\,s$^{-1}$, He~\Rmnum{1} at -5000 \,km\,s$^{-1}$, Fe~\Rmnum{2} at -5000 \,km\,s$^{-1}$, and Ca~\Rmnum{2} triplet at 6700 \,km\,s$^{-1}$.}
\label{fig:7}
\end{figure*}

\subsubsection{Evolution and comparison of photospheric-phase spectra}

Figure \ref{fig:7} shows the evolution and comparison of photospheric spectra of SN 2021dbg with those of other well-studied SNe \Rmnum{2}. The spectral evolution of SN 2021dbg is slower than that of most other SNe \Rmnum{2}, and the characteristic spectral lines appear later. In spectra during the first 48 days, H$\alpha$ and H$\beta$, are relatively wide with no obvious P-Cygni absorption profile, and the absorption of He~\Rmnum{1} $\lambda5876$~\AA, Ca~ \Rmnum{2} NIR triplet, and Fe~\Rmnum{2} $\lambda5169$\,\AA\ are also very shallow. These features are similar to those of SN 2014G \citep{2016MNRAS.462..137T}, SN 2015bf \citep{2021MNRAS.505.4890L}, and SN 2018hmx \citep{2019eeu..confE...7B}. 

After 48 days, H$\alpha$, H$\beta$, and $\lambda4686$\,\AA\  begin to appear with obvious P-Cygni  profiles, and the absorption components of He~\Rmnum{1} $\lambda5876$~\AA, Ca~\Rmnum{2} NIR triplet, and Fe~\Rmnum{2} $\lambda5169$\,\AA\  also begin to deepen, but the absorption depth is still shallower than that of a typical SNe \Rmnum{2}P such as SN 2017eaw \citep{2019ApJ...875..136V}. Instead, it is more similar to the line absorption of a typical SNe \Rmnum{2}L such as SN 2014G. The H$\alpha$ emission lines of SN 2021dbg, SN 2014G, SN 2015bf, and SN 2018hmx show a large extension on the blue side, and the peaks display a significant blueshift that exceeds 3000\,km\,s$^{-1}$ at early phases and gradually decreases with time. SN 2017eaw also exhibits such a blueshift, but it is much smaller and decreases rapidly. This behavior has been reported in many SNe \Rmnum{2} \citep{1976ApJ...207..872C,2009MNRAS.397..677T}. \cite{2014ApJ...786...67A} proposed that this behavior is a direct consequence of a steep density profile in the ejecta, which translates into more confined line emission and higher occultation of the receding part of the ejecta. The rate of decrease of SN 2021dbg’s H$\alpha$ peak blueshift is slower than that of SN 2014G, SN 2015bf, and SN 2018hmx, which indicates that SN 2021dbg may have a thicker ejecta layer with a steep drop in matter density, electron density, and temperature.

\subsection{Bolometric luminosity analysis}
\subsubsection{Construction of bolometric luminosity curve}
\begin{figure*}
\plotone{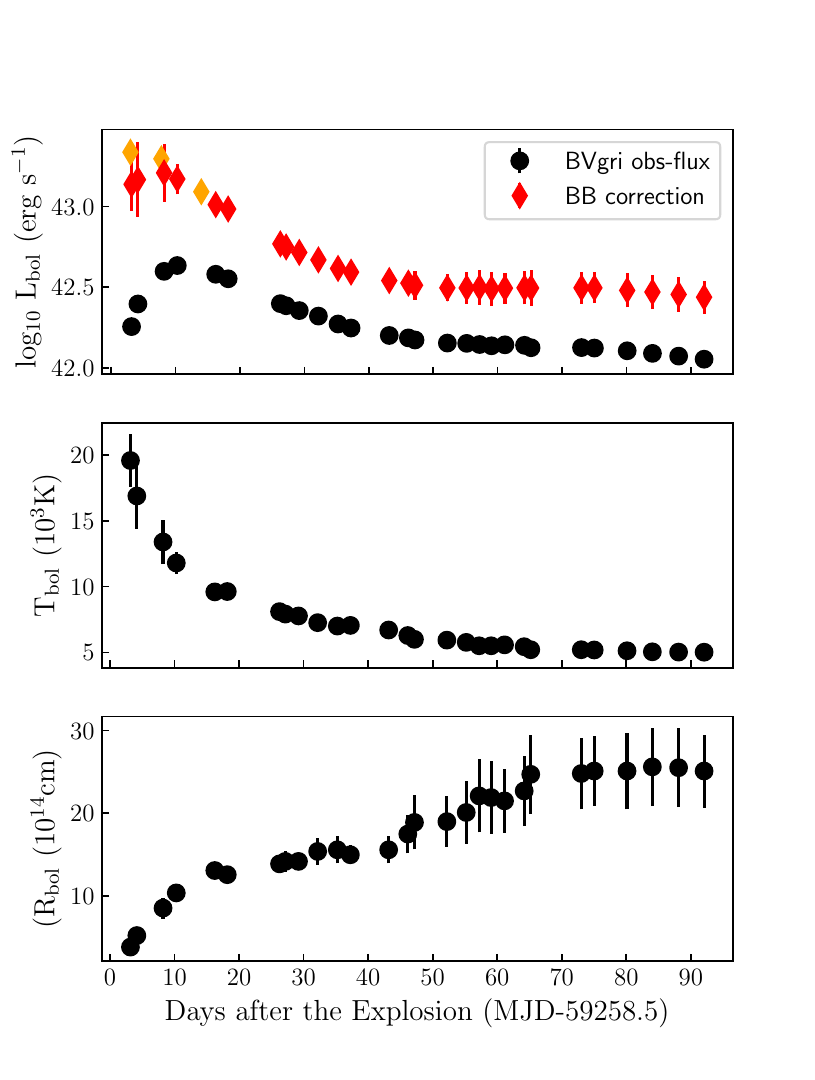}
\caption{Bolometric light curves of SN 2021dbg encompassing the pseudo-bolometric light curve derived from observed $BVgri$ photometry. Additionally, a bolometric light curve corrected through black body fitting, utilizing the same $BVgri$ photometry, has been included. Notably, three orange diamond-shaped points represent the UV flux-corrected luminosity, incorporating observations from {\it Swift}/UVOT. The photospheric temperature and radius computed by the SuperBol program are also showcased.}
\label{fig:8}
\end{figure*}

Within 100 days after the explosion, we obtained good optical photometric data and applied the SuperBol\footnote{\url{https://github.com/mnicholl/superbol}} program \citep{2018RNAAS...2..230N} to calculate the pseudo-bolometric luminosity, temperature, and radius evolution curve in the photosphere phase of SN 2021dbg, as shown in Figure \ref{fig:8}. In the UV bands, only three points were obtained by {\it Swift}/UVOT. We used them to correct the luminosity at early times, as indicated by the three orange diamond points in Figure \ref{fig:8}. Owing to the observation conditions, the photometric data obtained at late times are very limited, and the pseudo-bolometric luminosity cannot be obtained through multiband photometric data. 

\cite{2016MNRAS.459.3939V} successfully constructed the pseudo-bolometric luminosity curve from the beginning of radioactive decay to late times of 30 SNe \Rmnum{2} only by using the $V$ band. At late times, we obtained the most data in the $r$ band, and the evolutionary trend of the $r$-band light curve and the pseudo-bolometric luminosity curve are the most similar during the plateau period. Therefore, we applied the method of \citet{2016MNRAS.459.3939V} in the $r$ band and multiplied the $r$-band light curve by a coefficient to make it match the pseudo-bolometric luminosity curve during the plateau phase. Then, the pseudo-bolometric luminosity curve of SN 2021dbg in the radioactive-decay phase was effectively constructed by replacing the pseudo-bolometric luminosity curve with the $r$-band light curve, as shown with the red curve in Figure \ref{fig:11}.

\subsubsection{Evolution and comparison of bolometric curve}
In Figure \ref{fig:8}, SN 2021dbg reaches its peak luminosity, $L_{\rm max}= (1.7\pm0.9)\times 10^{43}$\,erg\,s$^{-1}$, at $\sim 8$ days after the explosion in the bolometric light curve derived from the observed $BVgri$ photometry. We noted that the spectral energy distribution derived from the $BVgri$ photometry at  early times was not adequately aligned with the blackbody fitting, resulting in significant errors and potential underestimation of the actual value. To address this issue, we utilized the UV data from {\it Swift} in combination with the optical observations for blackbody fitting. The resulting bolometric luminosity value of $L_{\rm max}= (2.1\pm0.4)\times 10^{43}$\,erg\,s$^{-1}$ is significantly higher than that obtained from the optical bands alone, and the associated uncertainty is reduced.

\begin{figure*}
\plotone{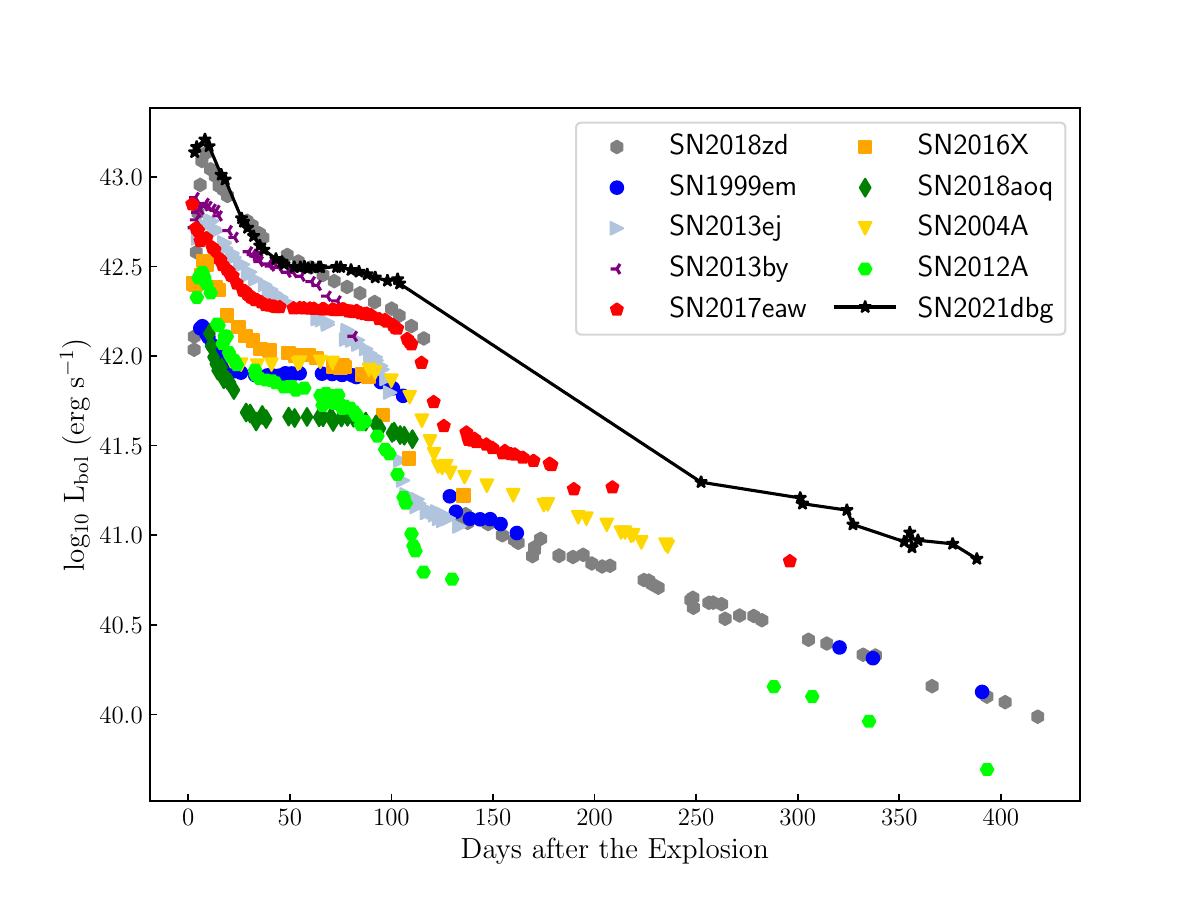}
\caption{Pseudobolometric light curve comparion of SN 2021dbg with that of some well-studied SNe \Rmnum{2}.}
\label{fig:9}
\end{figure*}

Figure \ref{fig:9} shows the pseudo-bolometric curve comparison of SN 2021dbg with other SNe \Rmnum{2}. The evolution trend of the bolometric luminosity curve of SN 2021dbg is similar to that of SN 1999em, SN 2016X, SN 2017eaw, and other typical SNe \Rmnum{2}P, but SN 2021dbg is brighter;  the peak bolometric luminosity of SN 2021dbg is about three times higher than that of SN 2017eaw, five times higher than that of SN 2016X, and about ten times higher than that of SN 1999em. Peaking in the near-UV at the same time as the optical light helps produce the high bolometric luminosity. Since SN 2017eaw showed early and moderate CSM interaction \citep{2019ApJ...876...19S,2019ApJ...875..136V}, SN 2021dbg with higher luminosity may have had even stronger early-time CSM interaction, or  higher total core energy released by the explosion. 

At $t < 50$\,d, SN 2018zd exhibits the closest luminosity to SN 2021dbg. However, a notable difference lies in the slower decline of SN 2018zd compared to SN 2021dbg, without a pronounced plateau beyond 50 days. At 120 days after the explosion, the bolometric luminosity of SN 2018zd nearly matches that of SN 1999em. Based on this observation, \cite{2020MNRAS.498...84Z} postulated that the enhanced luminosity of SN 2018zd relative to SN 1999em within the first 120 days results from the interaction between ejecta and CSM, while the explosion core energy of SN 2018zd remains comparable to that of SN 1999em. Furthermore, the pseudo-bolometric curve of SN 2021dbg displays a more prominent radioactive decay tail than do SN 2017eaw, SN 2018zd, and SN 1999em. This suggests a higher production of $^{56}$Ni in SN 2021dbg, as further elaborated in Section \ref{Sect 4.2}.

\section{Discussion}

\subsection{Ejecta Velocities}

\begin{figure*}
\gridline{
  \fig{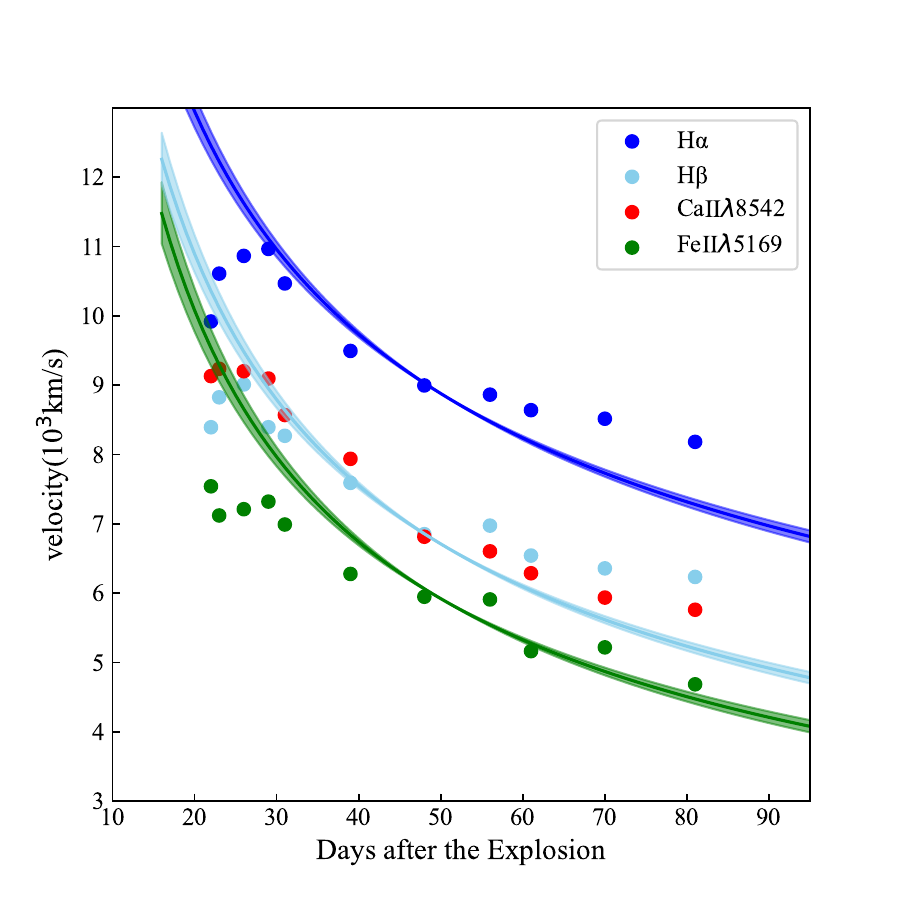}{0.48\textwidth}{(a)}
  \fig{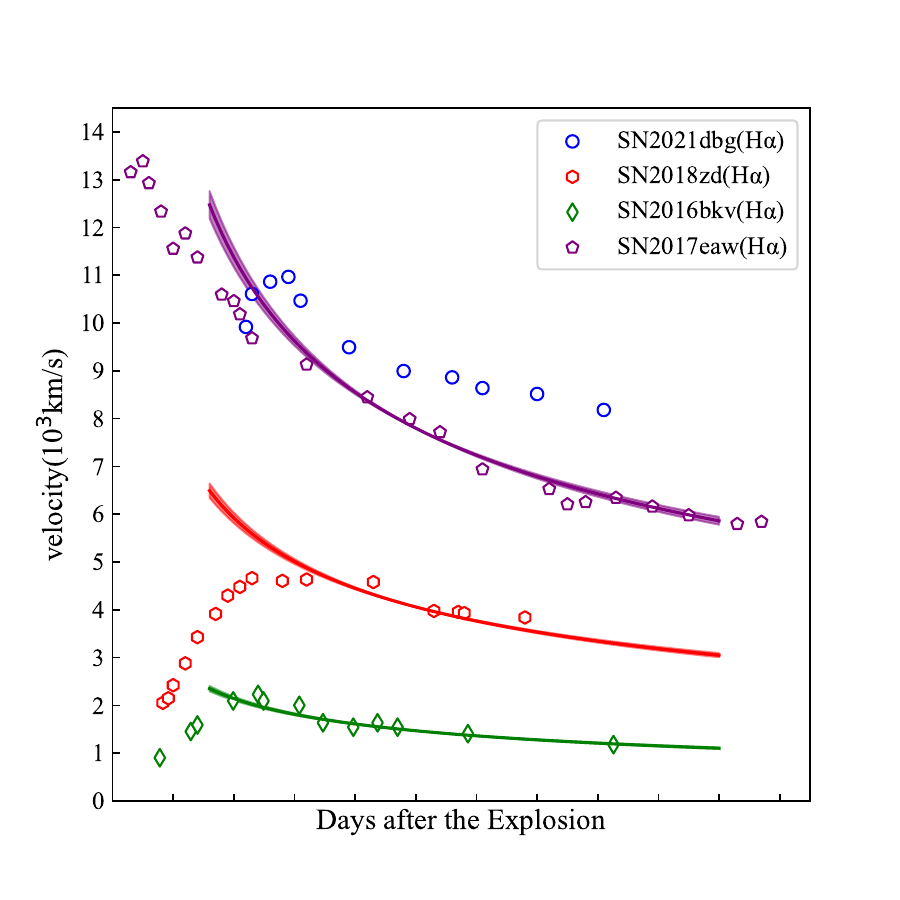}{0.48\textwidth}{(b)}
         }
\gridline{
  \fig{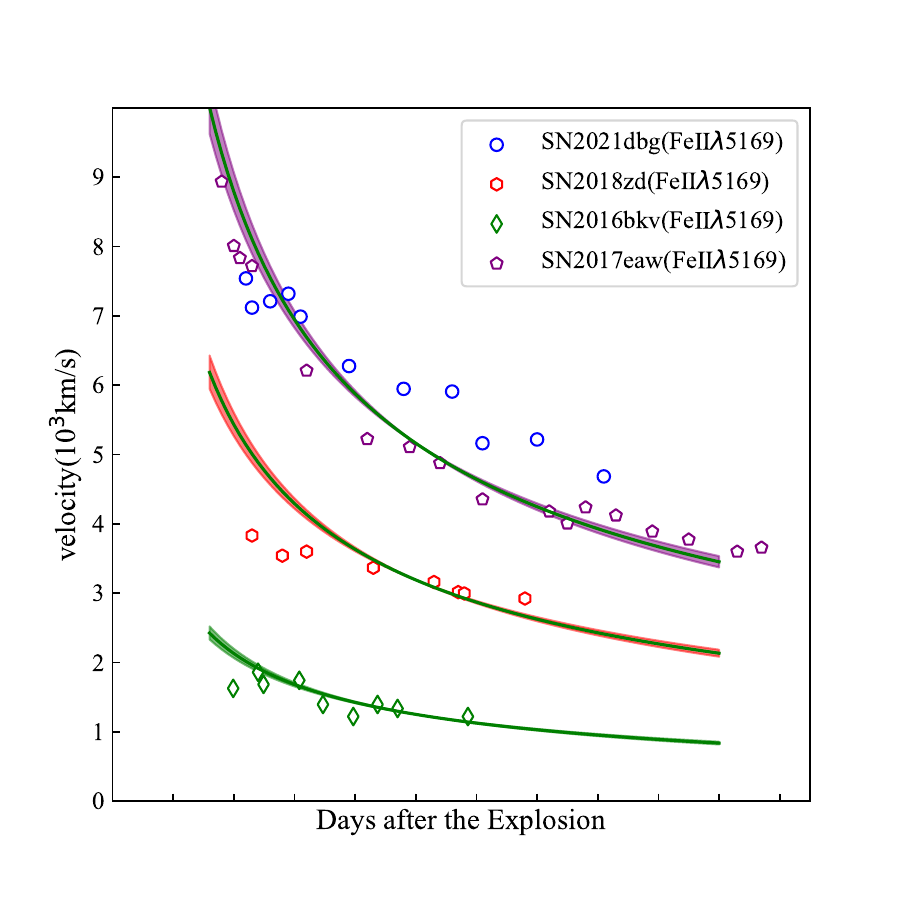}{0.48\textwidth}{(c)}
  \fig{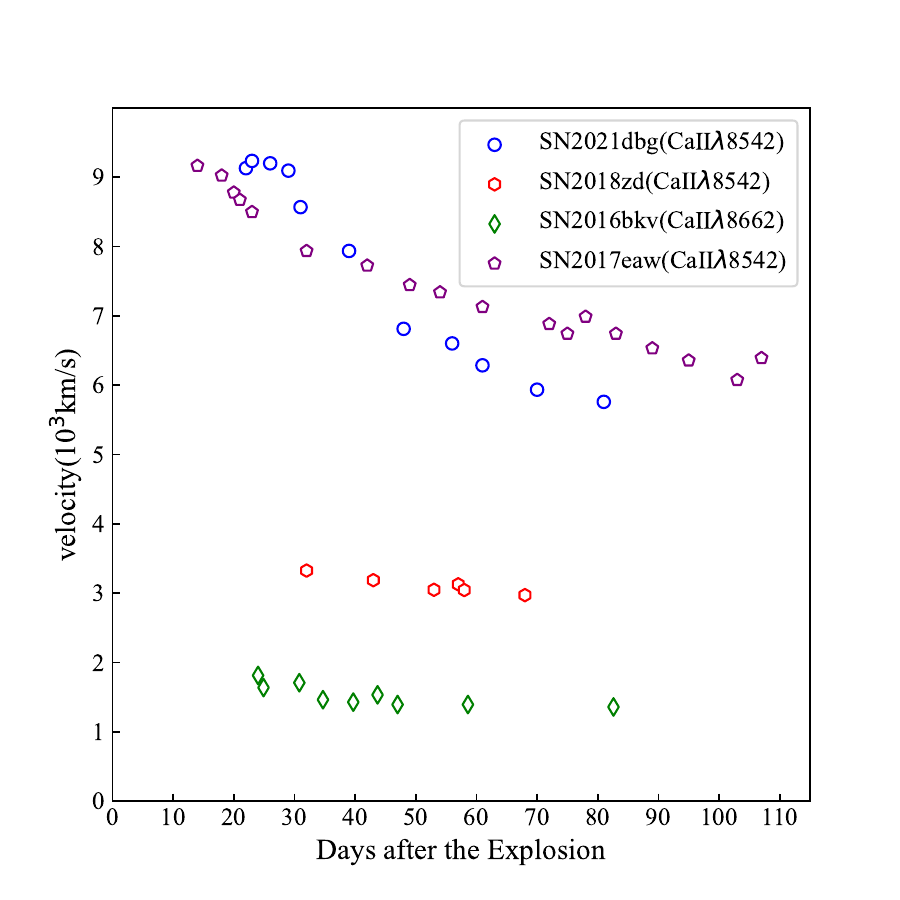}{0.48\textwidth}{(d)}
         }
\caption{{\it Panel (a):} SN 2021dbg's Velocity of ejecta calculated with H$\alpha$, H$\beta$, Ca~\Rmnum{2}$\lambda8542$\,\AA, and Fe~\Rmnum{2} $\lambda5169$\,\AA. The blue shaded lines represent the power-law fit for H$\alpha$ velocity; The skyblue shaded lines represent the power-law fit for H$\beta$ velocity; The green shaded lines represent the power-law fit for Fe~\Rmnum{2} $\lambda5169$\,\AA velocity.
{\it Panel (b):} Comparison of H$\alpha$ velocity between SN 2021dbg with other SNe \Rmnum{2}. The purple shaded lines represent the power-law fit for SN 2017eaw's H$\alpha$ velocity;
The red shaded lines represent the power-law fit for SN 2018zd's H$\alpha$ velocity; The green shaded lines represent the power-law fit for SN 2016bkv's H$\alpha$ velocity.
{\it Panel (c):} Comparison of Fe~\Rmnum{2} $\lambda5169$\,\AA velocity between SN 2021dbg with other SNe \Rmnum{2}. The purple shaded lines represent the power-law fit for SN 2017eaw's Fe~\Rmnum{2} $\lambda5169$\,\AA velocity;
The red shaded lines represent the power-law fit for SN 2018zd's Fe~\Rmnum{2} $\lambda5169$\,\AA velocity; The green shaded lines represent the power-law fit for SN 2016bkv's Fe~\Rmnum{2} $\lambda5169$\,\AA velocity.
{\it Panel (d):} Comparison of Ca~\Rmnum{2}$\lambda8542$\,\AA velocity between SN 2021dbg with other SNe \Rmnum{2}. 
These power-law fit methods for SNe \Rmnum{2}P ejecta velocity are provided by \cite{2014MNRAS.445..554F}}.
\label{fig:10}
\end{figure*}

The velocities of H$\alpha$, H$\beta$, Fe~\Rmnum{2} $\lambda5169$~\AA, and Ca \Rmnum{2} $\lambda8542$\,\AA\ are measured according to the blueshift of the wavelength of the minimum of the spectral line absorption profile relative to the rest wavelength of the emission line.  Minima of H$\alpha$, H$\beta$, and Fe~\Rmnum{2} $\lambda5169$\,\AA\ are obtained by polynomial fitting a P-Cygni profile. The minimum of Ca \Rmnum{2} $\lambda8542$\,\AA\ is obtained by fitting the absorption profile of the Ca\Rmnum{2} NIR triplet with three Gaussian components. Figure \ref{fig:11} presents the ejecta velocity evolution of SN 2021dbg and compares it with that of SN 2017eaw, SN 2018zd, and SN 2016bkv. These three SNe were chosen as comparison objects because SN 2017eaw has a similar ejecta velocity to SN 2021dbg, while SN 2018zd and SN 2016bkv exhibit hydrogen acceleration behavior in their early stages, similar to what is observed in SN 2021dbg. In Figure \ref{fig:10}, those shaded lines are a power-law fitting of the ejecta velocities of SNe \Rmnum{2}P summarized by \cite{2014MNRAS.445..554F}. 

The velocities of H$\alpha$ and H$\beta$ show an acceleration at $t<29$\,d and conform to the power law at $29<t<48$\,d. At $t>48$\,d, the decline in velocities slows down significantly. However, the velocity of Fe \Rmnum{2} $\lambda5169$\,\AA\ roughly conforms to the power law at $t>48$\,d, although it is lower than predicted by the power law at $t<48$\,d. The velocity of H$\beta$ is almost equal to that of Ca \Rmnum{2} $\lambda 8542$\,\AA, while lower than the H$\alpha$ velocity by about 2000\,km\,s$^{-1}$. Therefore, we think H$\alpha$ came from the outer layer of the ejecta, while H$\beta$ came from the inner layer of the ejecta near the core.

The early H-acceleration behavior observed in SN 2021dbg exhibits similarities to that of SN 2018zd \citep{2020MNRAS.498...84Z} and SN 2016bkv \citep{2018ApJ...859...78N}. According to \cite{2020MNRAS.498...84Z}, this accelerated H emission originates from either the wind matter or CSM located above the optically thick photosphere. Initially, these ionized winds or CSM are propelled by the shocked ejecta, resulting in the observed spectral lines and acceleration. Meanwhile, well-developed P-Cygni profiles are shaped within the ejecta. The H$\alpha$ P-Cygni profiles of SN 2021dbg have a shallow absorption at $22 < t < 29$\,d,  the absorption of the H$\alpha$ P-Cygni profiles thereafter is significantly deepened, and the H velocity accords with the power law at $29 < t < 48$\,d. These findings align with the hypothesis put forth by \cite{2020MNRAS.498...84Z}.

SN 2018zd showed a significant heating and accelerating behavior of the photosphere within a week after the explosion, which was explained by the reverse-shock heating of the outer layer of the ejecta \citep{2020MNRAS.498...84Z}. SN 2021dbg does not have such a heating and accelerating process of the photosphere at  early phases. After the P-Cygni profiles develop well, SN 2016bkv, SN 2018zd, and SN 2017eaw have a power-law velocity evolution of H$\alpha$, while the H$\alpha$ velocity of SN 2021dbg  declines more slowly and deviates from a power law significantly. A possible explanation for this behavior is that the progenitor of SN 2021dbg may have a thicker H shell, and the shell is likely to be layered because of a steep drop in matter density, electron density, and temperature. As the outermost layers dissipate, deeper H layers become visible.

\subsection{Estimate of the $^{56}$Ni Mass from the Tail Luminosity}\label{Sect 4.2}
\begin{figure*}
\plotone{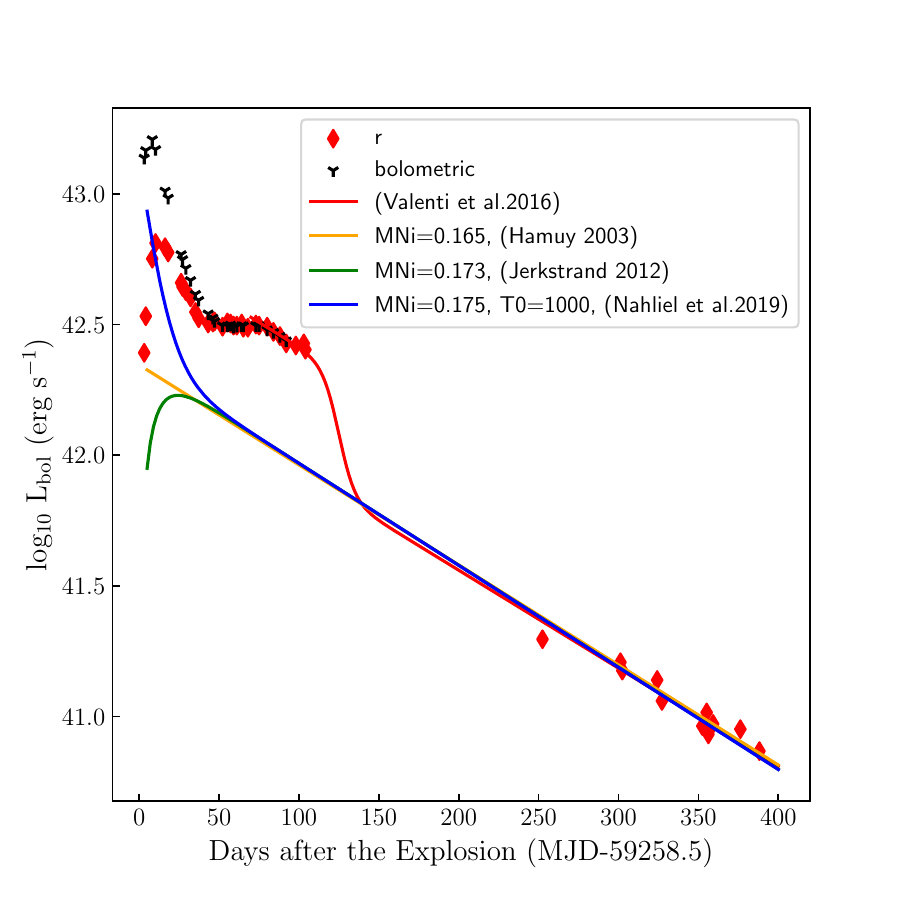}
\caption{Pseudobolometric light curve and $r$-band luminosity curve. The red curve is from the formula obtained by \cite{2016MNRAS.459.3939V}. Orange, green, and blue curves are the relationships between the tail bolometric luminosity and the $^{56}$Ni mass proposed by \cite{2003ApJ...582..905H}, \cite{2012A&A...546A..28J}, and \cite{2019MNRAS.484.3941W}, respectively.}
\label{fig:11}
\end{figure*}

In the late-time tail, the bolometric light curve is almost entirely powered by the radioactive decay of $^{56}$Co ($^{56}$Ni$\to$ $^{56}$Co$\to$ $^{56}$Fe), so the bolometric luminosity  is a good indication of the $^{56}$Ni mass synthesized in the explosion. Based on this theory, many researchers have given different methods or models to estimate the $^{56}$Ni mass, such as comparisons made with SN 1987A's tail luminosity \citep{2003MNRAS.338..939E}, the \cite{2003ApJ...582..905H} tail-luminosity method, and the \cite{2012A&A...546A..28J} tail-luminosity method. By comparing the tail luminosity with that of SN 1987A, the $^{56}$Ni mass of SN 2021dbg is found to be  0.17\,M$_{\odot}$. According to the methods of \cite{2003ApJ...582..905H} and \cite{2012A&A...546A..28J} (Eqs. (1) and (2) below, orange and green curves in Figure \ref{fig:11}, respectively), the mass of $^{56}$Ni is 0.165\,M$_{\odot}$ and 0.173\,M$_{\odot}$, respectively:

\begin{equation}
    L(t) = 1.271\times 10^{43}\frac{M_{\rm Ni}}{{M}_{\odot}}(exp[\frac{(t-t_{0})/(1+z)-6.1}{111.26}])^{-1}\,{erg}\,{s}^{-1}\, ,
\end{equation}

\begin{equation}
    L(t) = 9.92\times 10^{41}\frac{M_{\rm Ni}}{0.07\,{M}_{\odot}}(e^{-t/111.4}-e^{-t/8.8})\,{erg}\,{s}^{-1}\, .
\end{equation}

These above methods of calculating $^{56}$Ni from the tail luminosity assume that the energy released by radioactive decay was completely captured, the deposited energy was instantly re-emitted, and no other energy sources had any effect. Taking into account the case of $\gamma$-photon escape, \cite{2019MNRAS.484.3941W} added a parameter $T_{0}$ ($\gamma$-photon escape time) to the relationship between the energy of $^{56}$Co decay and the bolometric luminosity, constructing a new method for calculating the mass of $^{56}$Ni, as shown in Equations (3) and (4) below  from \cite{2023ApJS..269...40Y}:

\begin{equation}
    L(t)=M_{\rm Ni}((\epsilon_{\rm Ni}-\epsilon_{\rm Co})e^{-t/t_{\rm Ni}}+\epsilon_{\rm Co}e^{-t/t{Co}})(1-e^{-(t/T_{0})^{-2}})+0.034\,M_{\rm Ni}\epsilon_{\rm Co}(e^{-t/t_{\rm Ni}}-e^{-t/t_{\rm Co}})\, ,
\end{equation}

\begin{equation}
    T_{0}=\sqrt{C\kappa_{\gamma}M^{2}_{\rm ej}/E_{\rm kin}}\, .
\end{equation}
\noindent
where $\epsilon_{\rm Ni}=3.9 \times 10^{10}$\,erg\,g\,s$^{-1}$   and $\epsilon_{\rm Co}=6.8 \times 10^{9}$\,erg\,g\,s$^{-1}$  are the specific heating rates of Ni and Co decay, respectively, $t_{\rm Ni} = 8.8$ days and $t_{\rm Co} = 111.3$ days are their corresponding decay timescales, $M_{\rm ej}$ is the total mass of the ejecta, $E_{\rm kin}$ is the kinetic energy of the ejecta, $\kappa_{\gamma}$ is the gamma-ray opacity, and $C$ is a constant. The unit of $M_{\rm{Ni}}$ is gram.

Taking the matching tail luminosity as the standard, we set $T_{0} = 1000$ days and obtained a $^{56}$Ni mass of 0.175\,M$_{\odot}$, as seen in the blue curve in Figure \ref{fig:11}. Such a large value of $T_{0}$ indicates that almost no photons emitted by $^{56}$Ni decay escaped (as $T_{0} \to \infty$, Eq. (3) is equivalent to Eq. (2), meaning that $^{56}$Ni decay photons are completely captured), which implies that SN 2021dbg may have had a thick shell that almost completely captured the gamma-ray photons emitted during the decay of $^{56}$Ni. This is in agreement with the previous analysis of the spectra and ejecta velocities.

The values obtained from the above methods are all relatively close, indicating a mass of 56Ni in the range of 0.165 to 0.17\,M$_{\odot}$, which suggests that these methods are reasonable for SN 2021dbg. Since we do not have a more precise and reliable method to determine the mass of $^{56}$Ni, by combining the results of the aforementioned methods, we estimate the mass of $^{56}$Ni in SN 2021dbg as 0.17$\pm$ 0.05\,M$_{\odot}$.

\subsection{Modeling the Bolometric Light Curve and Explosion Parameters}\label{Sect 4.3}
We employ the radiation diffusion model, as outlined by \citet{1989ApJ...340..396A} and \citet{1989ApJ...340..414F}, to reconstruct the explosion parameters of SN 2021dbg. These calculations are facilitated by the LC2 code\footnote{Accessible at \url{https://titan.physx.u-szeged.hu/~nagyandi/LC2/}}, as detailed by \citet{2016A&A...589A..53N}. This model segments the ejecta resulting from the progenitor explosion into two distinct components: a high-mass, high-temperature, and high-density core enriched in He and heavier elements, encircled by a low-mass, low-temperature, and low-density shell predominantly composed of H. 

At early phases, the elevated luminosity primarily arises from the interaction between the shell and the CSM, while the thermal radiation emanating from the core predominantly contributes to the luminosity during the plateau. As the SN transitions into the tail phase, the luminosity is almost entirely sustained by the radioactive decay of $^{56}$Ni. Notably, the bolometric light curve of SN 2021dbg exhibits a prolonged linear decline at  early times, with a delayed plateau appearance. Attempts to fit the pseudo-bolometric light curve with a model featuring a core with a single shell proved unsuccessful.

Combining with the previous analysis, we considered that the progenitor of SN 2021dbg probably had a very thick H shell  whose matter density, electron density, and temperature probably dropped steeply. Hence, we tried to fit the pseudo-bolometric light curve with a core and two shells, achieving a successful result. This model finds that the outermost radius of the progenitor is $\sim 1200$\,R$_{\odot}$ and the total mass of the ejecta is 18.4\,M$_{\odot}$; considering a compact remnant, the mass of the progenitor is $\sim 20$\,M$_{\odot}$. 

\begin{figure*}
\plotone{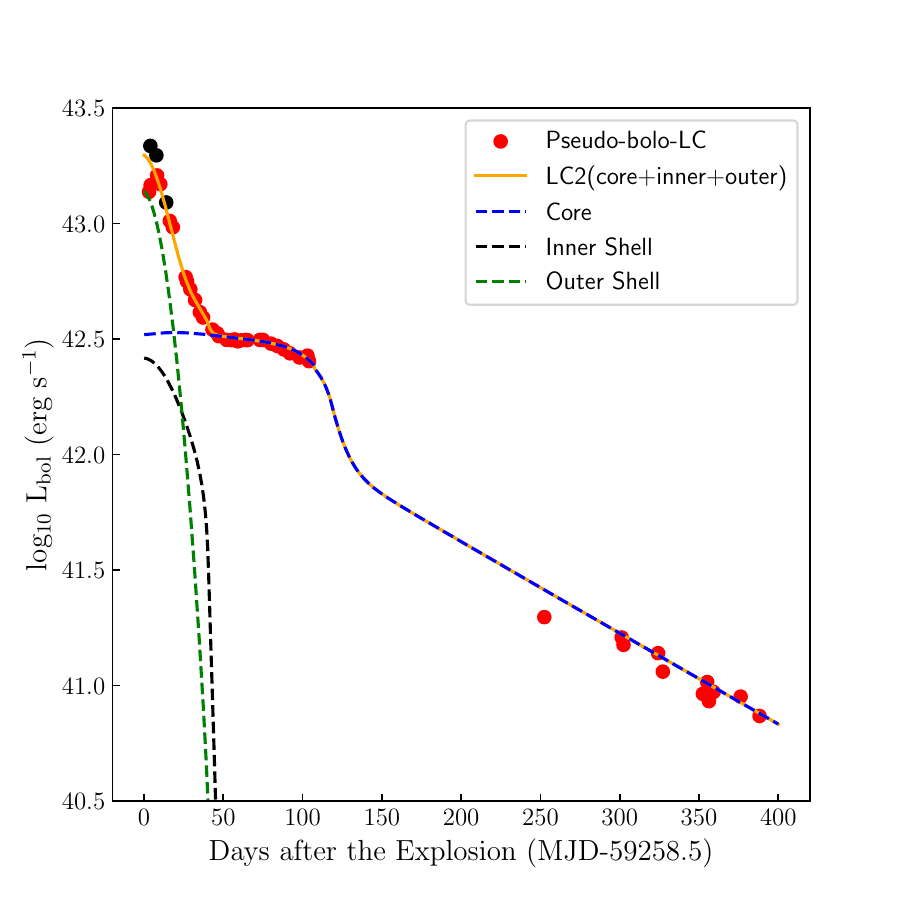}
\caption{Fitting result of the pseudobolometric luminosity curve. Three black points are corrected by {\it Swift} photometric data. Blue, black, and green curves are luminosity contributions from the core, inner shell, and outer shell, respectively; the orange curve is the total luminosity fitting to the pseudobolometric curve.}
\label{fig:12}
\end{figure*}

Figure \ref{fig:12} shows the result of fitting SN 2021dbg's pseudo-bolometric light curve, and the parameters of the model are presented in Table \ref{tab:2}. We divided the shell into inner and outer parts. The thickness of the inner shell is approximately equal to that of the outer shell, but the inner one has a higher mass, temperature, and Thomson scattering coefficient than the outer one. This means that the inner shell is denser and dominated primarily by ionized hydrogen, therefore resulting in a high electron density. The outer shell has a low density and is dominated primarily by neutral hydrogen. 

At early times following the explosion, the high-speed shell ejecta collide and interact with the CSM, accelerating the CSM and slowing down the shell ejecta. Much kinetic energy in the shell layer is converted into thermal radiation during this interaction, contributing to the early high luminosity. Most interactions last for $\sim 20$ days, and the luminosity decreases rapidly, then entering a plateau phase dominated by the core thermal radiation contribution. SN 2021dbg took $\sim 50$ days, a longer time than most SNe \Rmnum{2}, for the peak to decline to the plateau, likely owing to the thermal radiation contribution of the inner shell. The luminosity of the plateau and tail is mainly contributed by core thermal radiation and $^{56}$Ni decay.

Such a layered structure of the shell may be caused by unstable pulsations. This is likely due to the progenitor being in a pulsationally unstable state before the explosion and having entered or being on the verge of entering a phase of enhanced mass loss. Although it still retained a thick hydrogen-rich shell, the shell had already developed a layered structure due to unstable pulsations. \cite{2010ApJ...717L..62Y} provided simulation demonstrations showing that stars with initial masses between 17 to 20\,M$_{\odot}$ may experience enhanced mass loss in the form of pulsation-driven superwinds during the late stages of their evolution. Such pulsation-driven superwinds may be the key factor for the transition of supernova explosions from Type \Rmnum{2}P to Type \Rmnum{2}L.

\begin{deluxetable*}{cccccccc}
\tablenum{2}
\tablecaption{Parameters we used for the LC2 model.}
\label{tab:2}
\tablewidth{0pt}
\tablehead{
\nocolhead{} & \colhead{$M_{\rm ej}$ (M$_{\odot}$)} &\colhead{$R$ ($10^{13}$\,cm)} & \colhead{$T$ (K)} & \colhead{$E_{\rm kin}$ ($10^{51}$\,erg)} & \colhead{$E_{\rm th}$ ($10^{51}$\,erg)} & \colhead{$\kappa$ (cm$^{2}$\,g$^{-1}$)} & \colhead{$M_{\rm Ni}$ (M$_{\odot}$)}
}
\startdata
Core & 15.8 & 5.5 & 11000 & 1.7 & 1.1 & 0.24 & 0.165\\
Inner shell & 2.0 & 7.0 & 7000 & 0.91 & 0.13 & 0.36 & 0.0\\
Outer shell & 0.6 & 8.5 & 3000 & 0.075 & 0.085 & 0.18 & 0.0\\
\enddata
\end{deluxetable*}

\subsection{Mass Loss of the Progenitor }\label{Sect 4.4}
A large number of observational studies have shown that mass loss occurs at late times in massive stars, but it is still very difficult to directly observe the mass-loss process. The spectral flash emission lines at early phase are closely related to the CSM structure near the SN progenitor, and are thus helpful for us to infer the mass-loss process of the SN progenitor. The early spectra of SN 2021dbg show clear narrow H$\alpha$ emission, which allows us to estimate the mass-loss rate of its progenitor via the relationship between $L_{{\rm H}\alpha}$ and $\dot{\rm M}$ \citep{2013ApJ...768...47O}:
\begin{equation}
    L_{{\rm H}\alpha}\le\frac{4\pi h\nu_{{\rm H}\alpha}\alpha_{\rm eff}\beta K^{2}}{\langle\mu_{p}\rangle m_p^{2} r}\, ,
\end{equation}
\noindent
where $m_{p}$ is the mass of the proton, $\langle\mu_{p}\rangle$ is the mean number of nucleons per particle (mean molecular weight), $\beta\equiv(r_{1}-r)/r$, $K\equiv\dot{M}/(4\pi v_{w})$, $\alpha_{\rm eff} \approx 8.74 \times 10^{-14}(\frac{T_{\rm eff}}{10,000\,{\rm K}})^{0.89}$\,cm$^{3}$\,s$^{-1}$ \citep{2006agna.book.....O}. We have

\begin{equation}
    L_{{\rm H}\alpha}\le 3.68 \times 10^{39}\left(\frac{\dot{M}}{10^{-2}\,{\rm M}_{\odot}\,{\rm yr}^{-1}}\right)^{2}\left(\frac{v_{w}}{500\,{\rm km}\,{\rm s}^{-1}}\right)^{-2}\beta\left(\frac{r}{10^{15}\,{\rm cm}}\right)^{-1}\, {\rm erg}\,{\rm s}^{-1}\, .
\end{equation}
\noindent
Taking $T_{\rm eff} = 20,000$~K, which was obtained by blackbody fitting $\sim 3$ days after the explosion, Eq. (5) can be simplified to Eq. (6). We performed the Lorentzian function fitting of the H$\alpha$ line profile in the earliest spectrum obtained by NTT, including a single Lorentzian function, two Lorentzian functions, and three Lorentzian functions. 

The single Lorentzian profile fits the H$\alpha$ emission line well, indicating that the emission-line broadening is mainly caused by electron scattering rather than the velocity of the wind. The dual Lorentzian fitting aims to identify the narrow-line component within the emission line, but we cannot fit the effective narrow-line profile of the H$\alpha$ emission, probably limited by the low spectral resolution. The three-Lorentzian fitting is to consider the flux contribution of the emission lines of the possible host galaxy [N~\Rmnum{2}] $\lambda\lambda6548$,  6583\,\AA, and the results show that this flux contribution can be ignored. The luminosity of the single Lorentzian  fitting profile of the H$\alpha$ emission line is $L_{{\rm H}\alpha} = 2.1 \times 10^{39}$\,erg\,s$^{-1}$.

Taking the radius of the blackbody photosphere at the time of this earliest spectrum as $r \approx 3.8 \times 10^{14}$\,cm) and the radius of the blackbody photosphere at the beginning of the first plateau ($\sim 18$ days after the explosion; see Figure \ref{fig:8}) as $r_{1} \approx 1.3\times10^{15}$\,cm), because we think the first plateau indicates that the plateau phase of the photospheric radius around 18 day is attributed to the fact that the CSM beyond this distance is already very thin, and the collision between ejecta and CSM cannot generate sufficient radiation to expand the photosphere any further, then the photosphere gradually shifts to the inner shell. When the shell ejecta expanded to $r_{1}$, their temperature and density were too low to continue expanding the photosphere, until the hotter and denser core ejecta arrived and the photosphere continued expanding. This also explains the temperature exhibiting a slight bump as the photospheric radius expanded 40 days after the explosion. 

To estimate the mass loss rate, we also need to obtain the wind velocity. By fitting the H$\alpha$ flash emission line profile in the earliest spectrum, we obtain a velocity broadening of $\sim 1000$\,km\,s$^{-1}$. Such a wind velocity is unusually high for the progenitors of SNe \Rmnum{2}P and SNe \Rmnum{2}L, likely due to the fact that our first spectrum was obtained approximately 2.5 days after the explosion, when the wind material had already been accelerated by the ejecta. We compare an example with strong flash emission line and extremely early spectrum, SN 2013fs \citep{2017NatPh..13..510Y}. Assuming that SN 2021dbg and SN 2013fs have comparable wind speed ($\sim 100$\,km\,s$^{-1}$), we got a mass loss rate of $\dot{M} \ge 6 \times 10^{-4}$\,M$_{\odot}$\,yr$^{-1}$, slightly lower than the mass-loss rate of SN 2013fs ($\sim 10^{-3}$\,M$_{\odot}$\,yr$^{-1}$). According to Eq. (5) of \citet{2013ApJ...768...47O}, the H mass in the CSM is $\sim 1.7 \times 10^{-3}$\,M$_{\odot}$. If taking the typical wind speed of a red supergiant (RSG; $\sim 15$\,km\,s$^{-1}$) as $v_w$, we obtain $\dot{M} \ge 7 \times 10^{-6}$\,M$_{\odot}$\,yr$^{-1}$, which is a typical mass loss rate of RSGs.

The mass loss rate given by the model of \cite{2017A&A...605A..83D} is on the surface of the progenitor, while the narrow H$\alpha$ emission line we see comes from a farther place. The mass loss rate estimated by the H$\alpha$ flux is lower than the surface mass loss rate given by the model, reflecting the density change of CSM in the spatial scale. \cite{2024Natur.627..759Z} demonstrated that there may be a steep decline in the density of CSM within the distance of $\approx 5 \times 10^{13}$\,cm to $\approx 5 \times 10^{14}$\,cm.

\subsection{Masses Derived from the Nebular-Phase Spectrum}
\begin{figure*}
\centering \includegraphics[width=1.0\linewidth]{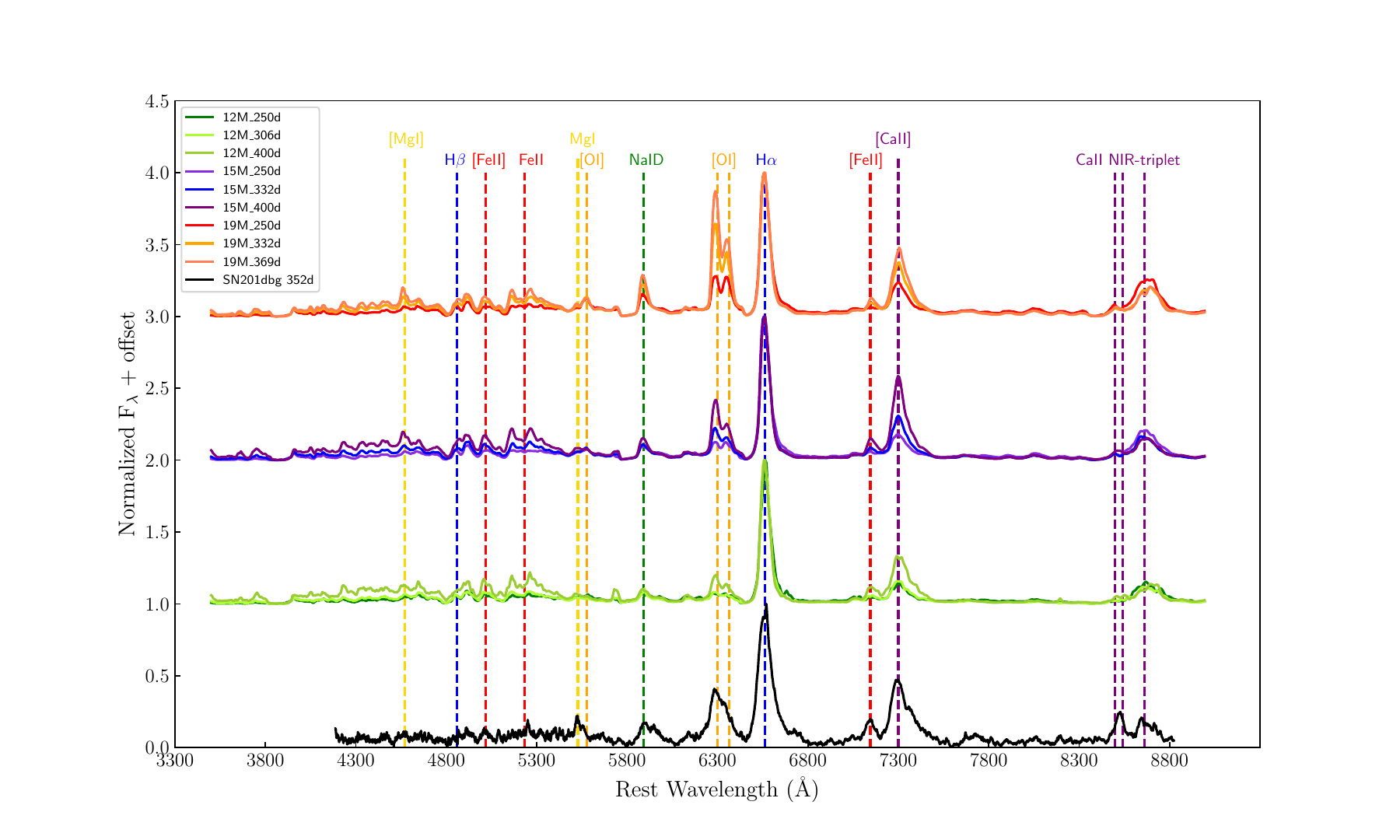}
\caption{Nebular-phase spectrum of SN 2021dbg obtained with Keck~1 compared with 12\,M$_{\odot}$, 15\,M$_{\odot}$, and 19\,M$_{\odot}$ model nebular spectra created by \cite{2012A&A...546A..28J}. All spectral fluxes have been normalized and vertically shifted.}
\label{fig:13}
\end{figure*}

\cite{2012A&A...546A..28J} proposed a model for estimating the mass of the progenitor of SNe from spectra in the nebular phase. As shown in Figure \ref{fig:13}, the nebular-phase spectrum, obtained $\sim 352$ days after the explosion using Keck~1, is compared with model spectra corresponding to various progenitor masses. The spectral lines, particularly, Na~\Rmnum{1}~D, Mg~\Rmnum{1}], and [O~\Rmnum{1}], are similar to the model spectra of the progenitor with a mass of 19\,M$_{\odot}$, implying that the progenitor of SN 2021dbg has a mass of $\sim 19$\,M$_{\odot}$.

In the nebular phase, \cite{2012A&A...546A..28J} also gave evolution curves of model spectral line luminosity ([O~\Rmnum{1}] $\lambda\lambda6300$,  6364\,\AA, Na~\Rmnum{1}~D, Mg~\Rmnum{
1}] $\lambda4571$\,\AA, H$\alpha$, [Fe~ \Rmnum{2}] $\lambda\lambda 7155$, 7172\,\AA, [Ca~\Rmnum{2}] $\lambda\lambda7291$, 7323\,\AA, Ca~\Rmnum{2} NIR triplet) for progenitors with different masses relative to the $^{56}$Co decay power over time. For the spectrum obtained at $t \approx 352$ days, the flux ratios  of the above spectral lines relative to the $^{56}$Co decay power are shown in Table \ref{tab:3}. The values of Mg~\Rmnum{1}] $\lambda4571$\,\AA\ and Na~\Rmnum{1}~D are $\sim 0.002$ higher, and the values of [Ca~\Rmnum{2}] $\lambda\lambda7291$, 7323\,\AA\ and Ca~\Rmnum{2} NIR triplet are $\sim 0.008$ higher, than the model values of the 19\,M$_{\odot}$ progenitor, while the ratios of [O~\Rmnum{1}] $\lambda\lambda 6300$, 6364\,\AA\ and H$\alpha$ are lower by $\sim 0.03$ and $\sim 0.05$ (respectively) than model values of the 19\,M$_{\odot}$ progenitor. The flux ratios of Na, Mg, and Ca are higher, and of H and O are lower, which means that the progenitor mass of SN 2021dbg may be greater than 19\,M$_{\odot}$.

Spectra from the nebular period can also be used to estimate the $^{56}$Ni mass produced in the ejecta. \cite{2003MNRAS.338..939E} proposed that the luminosity of the H$\alpha$ emission line in the nebular phase is proportional to the mass of $^{56}$Ni ejected, and \cite{2012MNRAS.420.3451M} obtained the relationship between the FWHM$_{\rm corr}$ (full width at half-maximum intensity that has been corrected for the instrumental broadening effect) of the nebular H$\alpha$ emission and the mass of $^{56}$Ni. We obtained  FWHM$_{\rm corr}\approx 92$\,\AA\ for the H$\alpha$ emission line in the nebular phase, and according to Eq. (7) below from \cite{2012MNRAS.420.3451M}, we  obtain $M_{\rm Ni}\approx 0.25_{-0.19}^{+0.70}\,{\rm M}_{\odot}$:

\begin{equation}
    M_{\rm Ni} = A \times 10^{B \times {\rm FWHM}_{\rm corr}}\,
    {\rm M}_{\odot}\, ,
\end{equation}
\noindent
where $A=1.8_{-0.7}^{+1.0} \times 10^{-3}$ and $B=0.023\pm0.004$.

\begin{figure*}
\plotone{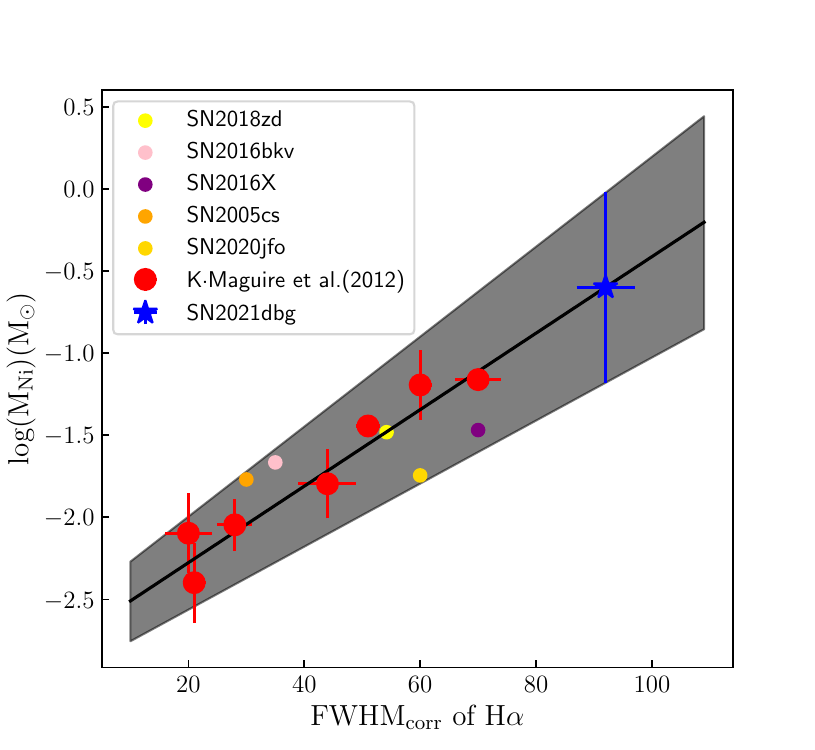}
\caption{The black line is the relationship obtained by \citet{2012MNRAS.420.3451M} between the FWHM$_{\rm corr}$ of the nebular H$\alpha$ emission and the mass of $^{56}$Ni. Shaded area is the uncertainty range.} 
\label{fig:14}
\end{figure*}

Figure \ref{fig:14} shows the correlation between the $^{56}$Ni mass estimated from the luminosity in the radioactive tail and the FWHM$_{\rm corr}$ of the nebular H$\alpha$ emission line, where the red data points are from \citet{2012MNRAS.420.3451M}, and the solid black line is the correlation obtained by \cite{2012MNRAS.420.3451M}. 
The uncertainty increases as FWHM$_{\rm corr}$ increases, and when FWHM$_{\rm corr}$ exceeds 80\,\AA, the uncertainty is too large. This is likely due to a lack of samples for high $^{56}$Ni mass and FWHM$_{\rm corr}$. Although the uncertainty is large, this result also supports a high $^{56}$Ni mass for SN 2021dbg.

\begin{deluxetable*}{cccccccc}
\tablenum{3}
\tablecaption{Ratio of the flux of spectral lines relative to $^{56}$Co decay power}
\label{tab:3}
\tablewidth{0pt}
\tablehead{
\colhead{Lines} & \colhead{H$\alpha$} &\colhead{[O~\Rmnum{1}]$  \lambda\lambda6300$, 6364} & \colhead{Na~\Rmnum{1}~D} & \colhead{Mg~\Rmnum{1}] $\lambda4571$} & \colhead{[Fe~\Rmnum{2}]$  \lambda\lambda7155$, 7172} & \colhead{[Ca~\Rmnum{2}]$  \lambda\lambda7291$, 7323} & \colhead{Ca~\Rmnum{2} NIR}
}
\startdata
Line F/$^{56}$Co power & 0.0534 & 0.0313 & 0.0128 & 0.0045 & 0.0143 & 0.0381 & 0.0267\\
\enddata
\end{deluxetable*}

\section{Conclusions}

We analyzed photometry of SN 2021dbg and found that the light-curve evolution is similar to that of SNe \Rmnum{2}P, but brighter overall than that of all typical SNe \Rmnum{2}P. Its luminosity is comparable to the middle brightness of SNe \Rmnum{2}L and slightly fainter than the median brightness of SNe \Rmnum{2}n. SN 2021dbg is bluer in color than most SNe \Rmnum{2}, because its temperature ($\sim 13,000$\,K at peak luminosity) and energy ($\sim 4.0\times10^{51}$\, erg, total kinetic energy and radiation energy) are higher than those of most SNe \Rmnum{2}. The energy released by SN 2021dbg is large, the ejecta velocity is high, and the spectral features of SN 2021dbg are closer to those of SNe \Rmnum{2}L, suggesting that SN 2021dbg is in the transition zone between SNe \Rmnum{2}P and SNe \Rmnum{2}L.

According to the tail luminosity, we estimate that SN 2021dbg produced  about $0.17\pm0.05$\,M$_{\odot}$ of $^{56}$Ni. However, such high $^{56}$Ni production is rare, according to statistics of \cite{2003ApJ...582..905H}; only SN 1992am (0.256\,M$_{\odot}$) and SN 1992af (0.156\,M$_{\odot}$) produced similar amounts of $^{56}$Ni. \cite{2019eeu..confE...7B} discovered another one, SN 2018hmx (0.14--0.17\,M$_{\odot}$), and proposed that these high-$^{56}$Ni-mass SNe \Rmnum{2} might link to 
high luminosities and velocities, but not to extreme progenitor parameters, and low metallicity might also play a role.

The early spectra exhibited flash emission lines, which disappeared within approximately 5 days, indicating that there was a small amount of CSM near the progenitor star of SN 2021dbg and  it was quickly wiped out by the ejecta. By analyzing the H$\alpha$ flash emission line, we estimated that the mass loss rate of the progenitor star was $1.0\times10^{-5}$ to $6.0\times10^{-4}$\,M$_{\odot}$\,yr$^{-1}$, and the mass of the CSM was about (1.0--2.0) $\times 10^{-3}$\,M$_{\odot}$, which had been produced about within 2--3\,yr before the explosion. This result shows that the progenitor of SN 2021dbg has an enhanced mass loss before the explosion.

The H$\alpha$ emission line of SN 2021dbg has a large blue-end broadening and a substantial blueshift at the peak in the photosphere phase, which is interpreted as a direct consequence of a steep density profile of the thick ejecta layer. This peak blueshift decreases at a relatively slower rate, suggesting that the progenitor of SN 2021dbg has a thick hydrogen-rich shell with a steep density profile. 

The pseudo-bolometric light curve is well-fitted by the method of the LC2 model with a core plus two shells. The model gives the progenitor a mass of $\sim 20$\,M$_{\odot}$ and an outermost radius of $\sim 1200$\,R$_{\odot}$, supporting that the progenitor is an RSG. It also suggests that the progenitor had a thick H shell of about 400\,R$_{\odot}$, and the thick H shell is approximately divided into two layers of equal thickness.  

The matter density, electron density, and temperature of the outer shell are significantly lower than those of the inner shell, suggesting that steep drops for these parameters exist in the transition region between the inner  and outer shells. Such a layered structure of the shell may be caused by unstable pulsations. 

Following the model proposed by \cite{2012A&A...546A..28J}, we estimated the mass of the progenitor star to be slightly greater than 19\,M$_{\odot}$ by using the nebular-phase spectrum taken with Keck~1 at $\sim 352$ days, in agreement with the value given by the LC2 model.

\begin{acknowledgments}
An anonymous referee made many useful suggestions that improved this paper. This work is supported by the National Key R\&D Program of China with grant 2021YFA1600404, the National Natural Science Foundation of China (12173082), the International Centre of Supernovae, Yunnan Key Laboratory (grant 202302AN360001), the science research grants from the China Manned Space Project (CMS-CSST-2021-A12), the Yunnan Fundamental Research Projects (grants 202401BC070007 and 202201AT070069), the Top-notch Young Talents Program of Yunnan Province, and the Light of West China Program provided by the Chinese Academy of Sciences. Y.-Z. Cai is supported by the National Natural Science Foundation of China (NSFC, grant 12303054) and the Yunnan Fundamental Research Projects (grant 202401AU070063).
The work of X.W. is supported by the National Natural Science Foundation of China (NSFC grants 12033003 and 12288102) and the Tencent Xplorer Prize.

A.V.F.’s research group at UC Berkeley acknowledges financial assistance from the Christopher R. Redlich Fund, Gary and Cynthia Bengier, Clark and Sharon Winslow, Alan Eustace, William Draper, Timothy and Melissa Draper, Briggs and Kathleen Wood, Sanford Robertson (W.Z. is a Bengier-Winslow-Eustace Specialist in Astronomy, T.G.B. is a Draper-Wood-Robertson Specialist in Astronomy, Y.Y. was a Bengier-Winslow-Robertson Fellow in Astronomy), and numerous other donors. 

Some of the data presented herein were obtained at the W. M. Keck
Observatory, which is operated as a scientific partnership among the
California Institute of Technology, the University of California, and
NASA; the observatory was made possible by the generous financial
support of the W. M. Keck Foundation.
We acknowledge the support of the staff of the LJT, XLT, and Keck Observatory. Funding for the LJT has been provided by the CAS and the People's Government of Yunnan Province. The LJT is jointly operated and administrated by YNAO and the Center for Astronomical Mega-Science, CAS.

\end{acknowledgments}

\bibliography{ms}{}
\bibliographystyle{aasjournal}

\begin{appendices}

\section{Photometric and Spectroscopic Data}
\renewcommand{\thefigure}{A\arabic{figure}} 
\setcounter{figure}{0}  

\begin{figure}[h]
\centering \includegraphics[width=0.6\linewidth]{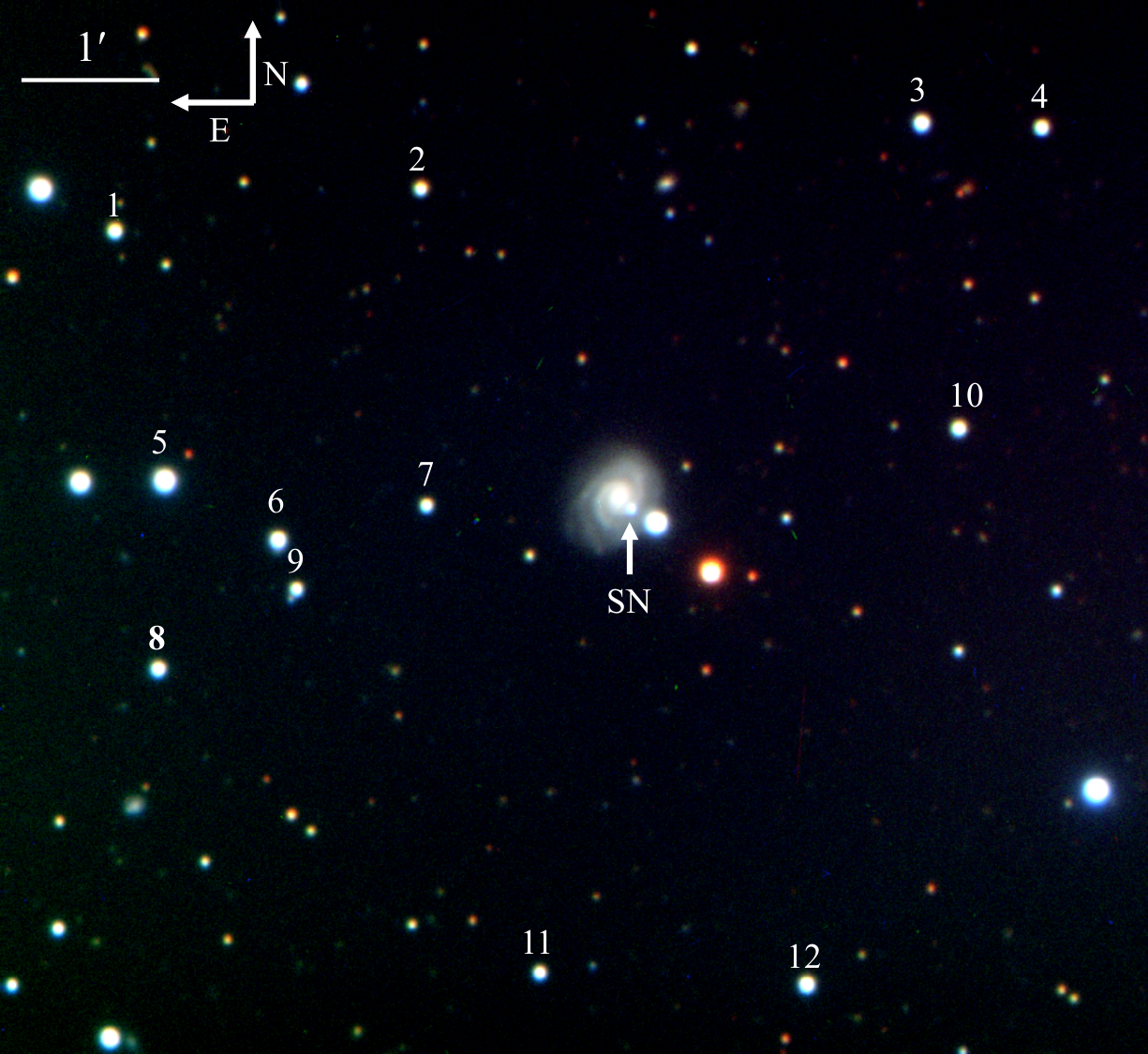}
\caption{Finder chart of SN 2021dbg. Information for these local reference stars is listed in Table \ref{tab:A2}.}
\label{fig:A1}
\end{figure}

\makeatletter  
\renewcommand{\thetable}{A\arabic{table}}  
\makeatother  
 
\setcounter{table}{0}
\begin{longtable}{cccccc} 
\caption[A.1]{Photometric data obtained by LJT, ATLAS and {\it Swift}.}\label{tab:A1}\\  
\toprule MJD & PHASE & TELESCOPE+INSTRUMENT & FILTER & MAG & EMAG \\    
\midrule    
\endfirsthead    
    
\multicolumn{6}{c}   
{{\bfseries Table \thetable{} -- Continued}} \\    
\toprule MJD & PHASE & TELESCOPE+INSTRUMENT & FILTER & MAG & EMAG \\    
\midrule    
\endhead    
    
\midrule    
\multicolumn{6}{c}{} \\    
\endfoot    
    
\bottomrule    
\endlastfoot  
59261.67 	&	3.17	&	LJT	&	B	&	17.04	&	0.01	\\
59262.66 	&	4.16	&	LJT	&	B	&	16.73	&	0.01	\\
59266.72 	&	8.22	&	LJT	&	B	&	16.34	&	0.01	\\
59268.77 	&	10.27	&	LJT	&	B	&	16.26	&	0.02	\\
59274.74 	&	16.24	&	LJT	&	B	&	16.5	&	0.02	\\
59276.67 	&	18.17	&	LJT	&	B	&	16.56	&	0.03	\\
59280.66 	&	22.16	&	LJT	&	B	&	16.87	&	0.01	\\
59284.77 	&	26.27	&	LJT	&	B	&	17.13	&	0.01	\\
59285.67 	&	27.17	&	LJT	&	B	&	17.19	&	0.01	\\
59287.71 	&	29.21	&	LJT	&	B	&	17.26	&	0.01	\\
59290.68 	&	32.18	&	LJT	&	B	&	17.46	&	0.01	\\
59293.73 	&	35.23	&	LJT	&	B	&	17.59	&	0.06	\\
59295.75 	&	37.25	&	LJT	&	B	&	17.64	&	0.04	\\
59301.68 	&	43.18	&	LJT	&	B	&	17.78	&	0.1	\\
59304.65 	&	46.15	&	LJT	&	B	&	17.95	&	0.1	\\
59305.68 	&	47.18	&	LJT	&	B	&	18.11	&	0.04	\\
59310.70 	&	52.2	&	LJT	&	B	&	18.15	&	0.03	\\
59313.70 	&	55.2	&	LJT	&	B	&	18.25	&	0.03	\\
59315.70 	&	57.2	&	LJT	&	B	&	18.35	&	0.02	\\
59317.56 	&	59.06	&	LJT	&	B	&	18.37	&	0.04	\\
59319.64 	&	61.14	&	LJT	&	B	&	18.31	&	0.02	\\
59322.68 	&	64.18	&	LJT	&	B	&	18.37	&	0.04	\\
59323.70 	&	65.2	&	LJT	&	B	&	18.49	&	0.04	\\
59331.53 	&	73.03	&	LJT	&	B	&	18.46	&	0.1	\\
59333.53 	&	75.03	&	LJT	&	B	&	18.47	&	0.06	\\
59338.62 	&	80.12	&	LJT	&	B	&	18.55	&	0.09	\\
59342.54 	&	84.04	&	LJT	&	B	&	18.62	&	0.05	\\
59346.61 	&	88.11	&	LJT	&	B	&	18.67	&	0.02	\\
59350.56 	&	92.06	&	LJT	&	B	&	18.71	&	0.02	\\
59364.54 	&	106.04	&	LJT	&	B	&	19.03	&	0.14	\\
59261.67	&	3.17	&	LJT	&	V	&	17.11	&	0.01	\\
59262.66	&	4.16	&	LJT	&	V	&	16.79	&	0.01	\\
59266.72	&	8.22	&	LJT	&	V	&	16.21	&	0.01	\\
59268.77	&	10.27	&	LJT	&	V	&	16.15	&	0.01	\\
59274.74	&	16.24	&	LJT	&	V	&	16.25	&	0.02	\\
59276.67	&	18.17	&	LJT	&	V	&	16.34	&	0.04	\\
59280.66	&	22.16	&	LJT	&	V	&	16.54	&	0.01	\\
59284.77	&	26.27	&	LJT	&	V	&	16.7	&	0.01	\\
59285.67	&	27.17	&	LJT	&	V	&	16.69	&	0.01	\\
59287.71	&	29.21	&	LJT	&	V	&	16.79	&	0.01	\\
59290.68	&	32.18	&	LJT	&	V	&	16.88	&	0.01	\\
59293.73	&	35.23	&	LJT	&	V	&	16.94	&	0.03	\\
59295.75	&	37.25	&	LJT	&	V	&	17.06	&	0.03	\\
59301.68	&	43.18	&	LJT	&	V	&	17.15	&	0.04	\\
59304.65	&	46.15	&	LJT	&	V	&	17.17	&	0.03	\\
59305.68	&	47.18	&	LJT	&	V	&	17.2	&	0.02	\\
59310.7	&	52.2	&	LJT	&	V	&	17.25	&	0.01	\\
59313.7	&	55.2	&	LJT	&	V	&	17.29	&	0.01	\\
59315.7	&	57.2	&	LJT	&	V	&	17.29	&	0.01	\\
59317.56	&	59.06	&	LJT	&	V	&	17.31	&	0.02	\\
59319.64	&	61.14	&	LJT	&	V	&	17.3	&	0.01	\\
59322.68	&	64.18	&	LJT	&	V	&	17.33	&	0.02	\\
59323.7	&	65.2	&	LJT	&	V	&	17.38	&	0.02	\\
59331.53	&	73.03	&	LJT	&	V	&	17.4	&	0.03	\\
59333.53	&	75.03	&	LJT	&	V	&	17.43	&	0.02	\\
59338.62	&	80.12	&	LJT	&	V	&	17.51	&	0.06	\\
59342.54	&	84.04	&	LJT	&	V	&	17.53	&	0.02	\\
59346.61	&	88.11	&	LJT	&	V	&	17.55	&	0.01	\\
59350.56	&	92.06	&	LJT	&	V	&	17.59	&	0.01	\\
59362.55	&	104.05	&	LJT	&	V	&	17.73	&	0.02	\\
59559.73	&	301.23	&	LJT	&	V	&	21.45	&	0.21	\\
59610.95	&	352.45	&	LJT	&	V	&	21.79	&	0.15	\\
59613.68	&	355.18	&	LJT	&	V	&	21.79	&	0.14	\\
59617.7	&	359.2	&	LJT	&	V	&	21.83	&	0.2	\\
59261.67	&	3.17	&	LJT	&	g	&	17.02	&	0.01	\\
59262.66	&	4.16	&	LJT	&	g	&	16.64	&	0.01	\\
59266.72	&	8.22	&	LJT	&	g	&	16.16	&	0.01	\\
59268.77	&	10.27	&	LJT	&	g	&	16.13	&	0.01	\\
59274.74	&	16.24	&	LJT	&	g	&	16.37	&	0.07	\\
59276.67	&	18.17	&	LJT	&	g	&	16.45	&	0.01	\\
59280.66	&	22.16	&	LJT	&	g	&	16.68	&	0.01	\\
59284.77	&	26.27	&	LJT	&	g	&	16.88	&	0.01	\\
59285.67	&	27.17	&	LJT	&	g	&	16.94	&	0.01	\\
59287.71	&	29.21	&	LJT	&	g	&	17.05	&	0.01	\\
59290.68	&	32.18	&	LJT	&	g	&	17.1	&	0.01	\\
59293.73	&	35.23	&	LJT	&	g	&	17.31	&	0.05	\\
59295.75	&	37.25	&	LJT	&	g	&	17.33	&	0.03	\\
59301.68	&	43.18	&	LJT	&	g	&	17.58	&	0.08	\\
59304.65	&	46.15	&	LJT	&	g	&	17.64	&	0.03	\\
59305.68	&	47.18	&	LJT	&	g	&	17.68	&	0.13	\\
59310.7	&	52.2	&	LJT	&	g	&	17.75	&	0.01	\\
59313.7	&	55.2	&	LJT	&	g	&	17.73	&	0.01	\\
59315.7	&	57.2	&	LJT	&	g	&	17.78	&	0.01	\\
59317.56	&	59.06	&	LJT	&	g	&	17.8	&	0.02	\\
59319.64	&	61.14	&	LJT	&	g	&	17.79	&	0.02	\\
59322.68	&	64.18	&	LJT	&	g	&	17.81	&	0.02	\\
59323.7	&	65.2	&	LJT	&	g	&	17.83	&	0.02	\\
59331.53	&	73.03	&	LJT	&	g	&	17.89	&	0.02	\\
59333.53	&	75.03	&	LJT	&	g	&	17.87	&	0.02	\\
59338.62	&	80.12	&	LJT	&	g	&	17.91	&	0.04	\\
59342.54	&	84.04	&	LJT	&	g	&	17.97	&	0.02	\\
59346.61	&	88.11	&	LJT	&	g	&	18.04	&	0.01	\\
59350.56	&	92.06	&	LJT	&	g	&	18.08	&	0.01	\\
59361.57	&	103.07	&	LJT	&	g	&	18.21	&	0.15	\\
59364.54	&	106.04	&	LJT	&	g	&	18.26	&	0.04	\\
59510.93	&	252.43	&	LJT	&	g	&	20.77	&	0.13	\\
59610.95	&	352.45	&	LJT	&	g	&	22.21	&	0.15	\\
59613.68	&	355.18	&	LJT	&	g	&	22.21	&	0.08	\\
59634.76	&	376.26	&	LJT	&	g	&	22.55	&	0.14	\\
59261.67	&	3.17	&	LJT	&	r	&	17.38	&	0.01	\\
59262.66	&	4.16	&	LJT	&	r	&	17.03	&	0.02	\\
59266.72	&	8.22	&	LJT	&	r	&	16.48	&	0.01	\\
59268.77	&	10.27	&	LJT	&	r	&	16.33	&	0.01	\\
59274.74	&	16.24	&	LJT	&	r	&	16.37	&	0.01	\\
59276.67	&	18.17	&	LJT	&	r	&	16.42	&	0.00	\\
59284.77	&	26.27	&	LJT	&	r	&	16.71	&	0.01	\\
59285.67	&	27.17	&	LJT	&	r	&	16.75	&	0.01	\\
59287.71	&	29.21	&	LJT	&	r	&	16.79	&	0.01	\\
59290.68	&	32.18	&	LJT	&	r	&	16.85	&	0.01	\\
59293.73	&	35.23	&	LJT	&	r	&	16.99	&	0.06	\\
59295.75	&	37.25	&	LJT	&	r	&	17.05	&	0.03	\\
59301.68	&	43.18	&	LJT	&	r	&	17.1	&	0.07	\\
59304.65	&	46.15	&	LJT	&	r	&	17.09	&	0.05	\\
59305.68	&	47.18	&	LJT	&	r	&	17.08	&	0.03	\\
59310.7	&	52.2	&	LJT	&	r	&	17.13	&	0.01	\\
59313.7	&	55.2	&	LJT	&	r	&	17.09	&	0.01	\\
59315.7	&	57.2	&	LJT	&	r	&	17.1	&	0.01	\\
59317.56	&	59.06	&	LJT	&	r	&	17.12	&	0.01	\\
59319.64	&	61.14	&	LJT	&	r	&	17.12	&	0.02	\\
59322.68	&	64.18	&	LJT	&	r	&	17.1	&	0.01	\\
59323.7	&	65.2	&	LJT	&	r	&	17.14	&	0.02	\\
59331.53	&	73.03	&	LJT	&	r	&	17.11	&	0.02	\\
59333.53	&	75.03	&	LJT	&	r	&	17.12	&	0.01	\\
59338.62	&	80.12	&	LJT	&	r	&	17.13	&	0.04	\\
59342.54	&	84.04	&	LJT	&	r	&	17.18	&	0.01	\\
59346.61	&	88.11	&	LJT	&	r	&	17.22	&	0.01	\\
59350.56	&	92.06	&	LJT	&	r	&	17.29	&	0.01	\\
59356.55	&	98.05	&	LJT	&	r	&	17.31	&	0.02	\\
59361.57	&	103.07	&	LJT	&	r	&	17.29	&	0.05	\\
59362.55	&	104.05	&	LJT	&	r	&	17.35	&	0.01	\\
59510.93	&	252.43	&	LJT	&	r	&	20.12	&	0.06	\\
59559.73	&	301.23	&	LJT	&	r	&	20.34	&	0.06	\\
59610.95	&	352.45	&	LJT	&	r	&	20.95	&	0.13	\\
59613.68	&	355.18	&	LJT	&	r	&	20.82	&	0.14	\\
59617.7	&	359.2	&	LJT	&	r	&	20.93	&	0.11	\\
59634.76	&	376.26	&	LJT	&	r	&	20.98	&	0.12	\\
59261.67	&	3.17	&	LJT	&	i	&	17.57	&	0.01	\\
59262.66	&	4.16	&	LJT	&	i	&	17.13	&	0.05	\\
59266.72	&	8.22	&	LJT	&	i	&	16.55	&	0.01	\\
59268.77	&	10.27	&	LJT	&	i	&	16.37	&	0.01	\\
59274.74	&	16.24	&	LJT	&	i	&	16.39	&	0.01	\\
59276.67	&	18.17	&	LJT	&	i	&	16.47	&	0.01	\\
59280.66	&	22.16	&	LJT	&	i	&	16.67	&	0.01	\\
59284.77	&	26.27	&	LJT	&	i	&	16.77	&	0.01	\\
59285.67	&	27.17	&	LJT	&	i	&	16.79	&	0.01	\\
59287.71	&	29.21	&	LJT	&	i	&	16.85	&	0.01	\\
59290.68	&	32.18	&	LJT	&	i	&	16.89	&	0.01	\\
59293.73	&	35.23	&	LJT	&	i	&	16.96	&	0.02	\\
59295.75	&	37.25	&	LJT	&	i	&	17.01	&	0.02	\\
59301.68	&	43.18	&	LJT	&	i	&	17.11	&	0.05	\\
59304.65	&	46.15	&	LJT	&	i	&	17.12	&	0.12	\\
59305.68	&	47.18	&	LJT	&	i	&	17.15	&	0.04	\\
59310.7	&	52.2	&	LJT	&	i	&	17.17	&	0.01	\\
59313.7	&	55.2	&	LJT	&	i	&	17.18	&	0.01	\\
59315.7	&	57.2	&	LJT	&	i	&	17.15	&	0.01	\\
59317.56	&	59.06	&	LJT	&	i	&	17.17	&	0.01	\\
59319.64	&	61.14	&	LJT	&	i	&	17.15	&	0.02	\\
59322.68	&	64.18	&	LJT	&	i	&	17.14	&	0.01	\\
59323.7	&	65.2	&	LJT	&	i	&	17.12	&	0.01	\\
59331.53	&	73.03	&	LJT	&	i	&	17.11	&	0.02	\\
59333.53	&	75.03	&	LJT	&	i	&	17.1	&	0.01	\\
59338.62	&	80.12	&	LJT	&	i	&	17.15	&	0.04	\\
59342.54	&	84.04	&	LJT	&	i	&	17.17	&	0.01	\\
59346.61	&	88.11	&	LJT	&	i	&	17.21	&	0.01	\\
59350.56	&	92.06	&	LJT	&	i	&	17.24	&	0.01	\\
59356.55	&	98.05	&	LJT	&	i	&	17.28	&	0.02	\\
59361.57	&	103.07	&	LJT	&	i	&	17.29	&	0.06	\\
59362.55	&	104.05	&	LJT	&	i	&	17.31	&	0.01	\\
59510.93	&	252.43	&	LJT	&	i	&	19.62	&	0.08	\\
59559.73	&	301.23	&	LJT	&	i	&	20.12	&	0.08	\\
59610.95	&	352.45	&	LJT	&	i	&	20.88	&	0.15	\\
59613.68	&	355.18	&	LJT	&	i	&	20.72	&	0.12	\\
59617.7	&	359.2	&	LJT	&	i	&	20.89	&	0.12	\\
59634.76	&	376.26	&	LJT	&	i	&	20.93	&	0.11	\\
59260.45 	&	1.95 	&	ATLAS-01	&	c	&	18.04	&	0.03	\\
59261.45 	&	2.95 	&	ATLAS-01	&	c	&	17.25	&	0.01	\\
59262.45 	&	3.95 	&	ATLAS-02	&	c	&	16.84	&	0.01	\\
59264.44 	&	5.94 	&	ATLAS-01	&	c	&	16.58	&	0.02	\\
59264.45 	&	5.95 	&	ATLAS-01	&	o	&	16.70 	&	0.01	\\
59266.45 	&	7.95 	&	ATLAS-02	&	o	&	16.42	&	0.01	\\
59268.38 	&	9.88 	&	ATLAS-01	&	o	&	16.34	&	0.01	\\
59270.50 	&	12.00 	&	ATLAS-01	&	c	&	16.35	&	0.01	\\
59276.41 	&	17.91 	&	ATLAS-01	&	o	&	16.45	&	0.01	\\
59276.55 	&	18.05 	&	ATLAS-01	&	c	&	16.45	&	0.01	\\
59278.46 	&	19.96 	&	ATLAS-02	&	o	&	16.49	&	0.01	\\
59280.46 	&	21.96 	&	ATLAS-01	&	c	&	16.58	&	0.01	\\
59280.40 	&	21.90 	&	ATLAS-01	&	o	&	16.53	&	0.01	\\
59288.36 	&	29.86 	&	ATLAS-01	&	o	&	16.80 	&	0.01	\\
59290.50 	&	32.00 	&	ATLAS-01	&	c	&	16.91	&	0.02	\\
59292.34 	&	33.84 	&	ATLAS-01	&	o	&	16.85	&	0.01	\\
59294.37 	&	35.87 	&	ATLAS-02	&	o	&	16.90 	&	0.02	\\
59300.38 	&	41.88 	&	ATLAS-01	&	o	&	17.02	&	0.05	\\
59300.50 	&	42.00 	&	ATLAS-01	&	c	&	17.20 	&	0.02	\\
59304.43 	&	45.93 	&	ATLAS-01	&	o	&	16.92	&	0.02	\\
59306.38 	&	47.88 	&	ATLAS-02	&	o	&	16.99	&	0.01	\\
59307.41 	&	48.91 	&	ATLAS-01	&	c	&	17.38	&	0.02	\\
59308.33 	&	49.83 	&	ATLAS-01	&	c	&	17.38	&	0.02	\\
59312.34 	&	53.84 	&	ATLAS-01	&	c	&	17.40 	&	0.02	\\
59322.32 	&	63.82 	&	ATLAS-02	&	o	&	17.07	&	0.02	\\
59324.31 	&	65.81 	&	ATLAS-01	&	o	&	17.08	&	0.02	\\
59326.29 	&	67.79 	&	ATLAS-02	&	o	&	17.01	&	0.02	\\
59328.33 	&	69.83 	&	ATLAS-01	&	o	&	17.08	&	0.03	\\
59331.34 	&	72.84 	&	ATLAS-01	&	o	&	17.06	&	0.02	\\
59332.29 	&	73.79 	&	ATLAS-01	&	o	&	17.11	&	0.02	\\
59334.30 	&	75.80 	&	ATLAS-02	&	o	&	17.10 	&	0.01	\\
59336.32 	&	77.82 	&	ATLAS-01	&	o	&	17.16	&	0.01	\\
59338.33 	&	79.83 	&	ATLAS-02	&	c	&	17.55	&	0.04	\\
59342.29 	&	83.79 	&	ATLAS-02	&	c	&	17.56	&	0.02	\\
59344.29 	&	85.79 	&	ATLAS-01	&	o	&	17.16	&	0.01	\\
59346.27 	&	87.77 	&	ATLAS-02	&	c	&	17.63	&	0.02	\\
59348.26 	&	89.76 	&	ATLAS-01	&	o	&	17.24	&	0.01	\\
59350.26 	&	91.76 	&	ATLAS-02	&	c	&	17.63	&	0.02	\\
59352.27 	&	93.77 	&	ATLAS-01	&	o	&	17.25	&	0.02	\\
59354.30 	&	95.80 	&	ATLAS-02	&	o	&	17.24	&	0.03	\\
59358.26 	&	99.76 	&	ATLAS-02	&	o	&	17.25	&	0.03	\\
59359.25 	&	100.75 	&	ATLAS-01	&	o	&	17.25	&	0.02	\\
59361.26 	&	102.76 	&	ATLAS-02	&	o	&	17.27	&	0.03	\\
59368.27 	&	109.77 	&	ATLAS-01	&	c	&	17.93	&	0.02	\\
59376.25 	&	117.75 	&	ATLAS-01	&	c	&	18.07	&	0.05	\\
59262.41	&	3.91	&	SWIFT+UVOT	&	uw2	&	14.64	&	0.05	\\
59266.28	&	7.78	&	SWIFT+UVOT	&	uw2	&	15.05	&	0.07	\\
59272.51	&	14.01	&	SWIFT+UVOT	&	uw2	&	16.3	&	0.08	\\
59266.28	&	7.78	&	SWIFT+UVOT	&	um2	&	14.81	&	0.07	\\
59272.51	&	14.01	&	SWIFT+UVOT	&	um2	&	16.22	&	0.08	\\
59262.41	&	3.91	&	SWIFT+UVOT	&	uw1	&	14.93	&	0.07	\\
59266.28	&	7.78	&	SWIFT+UVOT	&	uw1	&	14.87	&	0.08	\\
59272.51	&	14.01	&	SWIFT+UVOT	&	uw1	&	15.89	&	0.09	\\
59262.41	&	3.91	&	SWIFT+UVOT	&	u	&	15.34	&	0.06	\\
59266.28	&	7.78	&	SWIFT+UVOT	&	u	&	15.05	&	0.08	\\
59272.51	&	14.01	&	SWIFT+UVOT	&	u	&	15.42	&	0.07	\\
59262.41	&	3.91	&	SWIFT+UVOT	&	b	&	16.57	&	0.08	\\
59266.28	&	7.78	&	SWIFT+UVOT	&	b	&	16.15	&	0.09	\\
59272.51	&	14.01	&	SWIFT+UVOT	&	b	&	16.3	&	0.08	\\
59262.41	&	3.91	&	SWIFT+UVOT	&	v	&	16.39	&	0.09	\\
59266.28	&	7.78	&	SWIFT+UVOT	&	v	&	16.1	&	0.13	\\
59272.51	&	14.01	&	SWIFT+UVOT	&	v	&	16.06	&	0.11	\\
\end{longtable}
\footnotesize 
\noindent{N\scriptsize{OTE}\footnotesize---Vega magnitude is used in $BV$ bands and $Swift$-UVOT, AB magnitude is used in $gri$ bands and ATLAS $co$ bands.} 
\normalsize 
\vskip 0.5cm  

\begin{deluxetable*}{cccccccccccccc}[h]
\tablenum{A2}
\tablecaption{Reference star photometric data in the field of SN 2021dbg.}
\label{tab:A2}
\tablehead{
\colhead{Star} & \colhead{$\alpha$ (J2000)} &\colhead{$\delta$ (J2000)} & \colhead{$g$} & \colhead{$\delta g$} & \colhead{$r$} & \colhead{$\delta r$} & \colhead{$i$} & \colhead{$\delta i$} & \colhead{$B$} & \colhead{$\delta B$} & \colhead{$V$} & \colhead{$\delta V$} & \colhead{Pan-STARRS detection}
}
\startdata
1 & 09:24:41.79 & -06:32:52.45 & 16.90 & 0.02 & 16.08 & 0.01 & 15.77 & 0.01 & 17.40 & 0.02 & 16.41 & 0.01 & $\ge$12 \\
2 & 09:24:32.92 & -06:32:34.25 & 17.58 & 0.02 & 16.53 & 0.02 & 16.09 & 0.01 & 18.17 & 0.03 & 16.96 & 0.01 & $\ge$12 \\
3 & 09:24:18.34 & -06:32:04.92 & 16.08 & 0.01 & 15.66 & 0.01 & 15.53 & 0.02 & 16.44 & 0.01 & 15.83 & 0.01 & $\ge$9 \\
4 & 09:24:14.79 & -06:32:07.26 & 16.96 & 0.03 & 16.24 & 0.01 & 15.94 & 0.01 & 17.43 & 0.02 & 16.53 & 0.01 & $\ge$9 \\
5 & 09:24:40.38 & -06:34:41.14 & 14.31 & 0.01 & 14.07 & 0.01 & 13.99 & 0.01 & 14.61 & 0.01 & 14.16 & 0.00 & $\ge$10 \\
6 & 09:24:37.05 & -06:35:07.05 & 15.89 & 0.01 & 15.54 & 0.01 & 15.43 & 0.02 & 16.23 & 0.01 & 15.68 & 0.01 & $\ge$10 \\
7 & 09:24:32.72 & -06:34:50.41 & 17.24 & 0.02 & 16.69 & 0.01 & 16.49 & 0.01 & 17.65 & 0.01 & 16.91 & 0.01 & $\ge$12 \\
8 & 09:24:40.60 & -06:36:02.30 & 16.35 & 0.02 & 15.92 & 0.01 & 15.78 & 0.02 & 16.72 & 0.01 & 16.09 & 0.01 & $\ge$14 \\
9 & 09:24:36.58 & -06:35:27.53 & 17.12 & 0.02 & 16.63 & 0.01 & 16.47 & 0.01 & 17.51 & 0.01 & 16.83 & 0.01 & $\ge$10 \\
10 & 09:24:17.22 & -06:34:17.96 & 16.73 & 0.02 & 16.01 & 0.01 & 15.75 & 0.01 & 17.20 & 0.02 & 16.30 & 0.01 & $\ge$11 \\
11 & 09:24:29.44 & -06:38:14.73 & 17.07 & 0.02 & 16.64 & 0.01 & 16.47 & 0.02 & 17.44 & 0.01 & 16.81 & 0.01 & $\ge$9 \\
12 & 09:24:21.65 & -06:38:20.25 & 16.10 & 0.01 & 15.73 & 0.01 & 15.61 & 0.01 & 16.45 & 0.01 & 15.88 & 0.01 & $\ge$8 \\
\enddata

\tablecomments{Vega magnitude is used in $BV$ bands and AB magnitude is used in $gri$ bands.}
\end{deluxetable*}

\begin{deluxetable*}{cccccc}[h]
\tablenum{A3}
\tablecaption{Log of spectroscopic observations of SN 2021dbg.}
\label{tab:A3}
\tablehead{
\colhead{UT Date} & \colhead{Phase (d)} &\colhead{Telescope} & \colhead{Instrument} & \colhead{Range (\AA)} & \colhead{R$^b$}
}
\startdata
2021-02-16 $^a$ 	& 	2.5	& 	NTT 	& 	EFOSC2 	& 	3650-8800 	& 	370	 \\
2021-02-16 	& 	3 & 	LJT & 	YFOSC+G3 	& 	3505-8765 	& 	350	\\
2021-02-18 	& 	5 & 	LJT & 	YFOSC+G14	 &	 3810-7435	 &	580	 \\
2021-02-19 	& 	6 & 	LJT & 	YFOSC+G3 	& 	3490-8740 	& 	350	\\
2021-02-21 	& 	8 & 	LJT & 	YFOSC+G3 	& 	3485-8740 	& 	350	\\
2021-02-23 	& 	10 &	 LJT &	 YFOSC+G3	 &	 3505-8765	 &	350	 \\
2021-03-07 	& 	22 &	 LJT &	 YFOSC+G3	 &	 3505-8765	 &	350	 \\
2021-03-08 	& 	23 &	 XLT &	 BFOSC+G4	 &	 3875-8825	 &	350	 \\
2021-03-11 	& 	26 &	 LJT &	 YFOSC+G3	 &	 3500-8760	 &	350	 \\
2021-03-14 	& 	29 &	 LJT &	 YFOSC+G3	 &	 3505-8765	 &	350	 \\
2021-03-16 	& 	31 &	 LJT &	 YFOSC+G3	 &	 3500-8760	 &	350	 \\
2021-03-24 	& 	39 &	 LJT &	 YFOSC+G3	 &	 4000-8760	 &	350	 \\
2021-04-02 	& 	48 &	 LJT &	 YFOSC+G3	 &	 3600-8765	 &	350	 \\
2021-04-10 	& 	56 &	 LJT &	 YFOSC+G3	 &	 3700-8765	 &	350	 \\
2021-04-15 	& 	61 &	 LJT &	 YFOSC+G3	 &	 4000-8765	 &	350	 \\
2021-04-24 	& 	70 &	 LJT &	 YFOSC+G3	 &	 4300-8765	 &	350	 \\
2021-05-05 	& 	81 &	 LJT &	 YFOSC+G3	 &	 3800-8760	 &	350	 \\
2022-01-31 	& 	352	& 	Keck 1	 & LRIS	 &	 4270-9000	 &	1600	 \\
\enddata
\tablecomments{Phase is relative to the estimated explosion date.}
$^a${The classification spectrum obtained by ePESSTO team  \citep{2021TNSCR.489....1H}.}\\
$^b${Spectral resolution defined as $\lambda/\Delta \lambda$.}
\end{deluxetable*}

\end{appendices}

\end{CJK}
\end{document}